\documentclass[letterpaper,conference]{IEEEtran}
\usepackage{cite}
\usepackage{amsmath,amssymb,amsfonts}
\usepackage{algorithmic}
\usepackage{graphicx}
\usepackage{textcomp}
\usepackage{subfigure}
\usepackage{epstopdf}
\usepackage{multicol}
\usepackage{multirow}
\usepackage{siunitx}
\usepackage{float}
\usepackage{footnote}
\usepackage[margin = 0.625in,top=0.75in,bottom=1in]{geometry}
\usepackage{draftwatermark}
\SetWatermarkText{Camera Ready}
\SetWatermarkScale{0.5}
\def\BibTeX{{\rm B\kern-.05em{\sc i\kern-.025em b}\kern-.08em
		T\kern-.1667em\lower.7ex\hbox{E}\kern-.125emX}}

\DeclareMathOperator*{\argmax}{arg\,max}
\newcounter{mytempeqncnt}
\begin{document}
	
	\title{Energy Efficient Event Localization and Classification for Nano IoT
		\vspace{-4.5mm}
	}
	\author{\IEEEauthorblockN{Shree Prasad M., Trilochan Panigrahi}
		\IEEEauthorblockA{\textit{Dept. of Electronics and Communication Engineering.} \\
			\textit{National Institute of Technology Goa},
			Goa, India \\
			shreeprasadm@gmail.com, tpanigrahi@nitgoa.ac.in}
		\vspace{-11.1mm}
		\and
		\IEEEauthorblockN{Mahbub Hassan}
		\IEEEauthorblockA{\textit{School of Computer Science and Engineering} \\
			\textit{University of New South Wales},
			Sydney, Australia \\
			mahbub@cse.unsw.edu.au}
		\vspace{-11.1mm}
	}
	\maketitle
	\begin{abstract}
		Advancements in nanotechnology promises new capabilities for Internet of Things (IoT) to monitor extremely fine-grained events by deploying sensors as small as a few hundred nanometers. Researchers predict that such tiny sensors can transmit wireless data using graphene-based nano-antenna radiating in the terahertz band (0.1-10 THz). Powering such wireless communications with nanoscale energy supply, however, is a major challenge to overcome. In this paper, we propose an energy efficient event monitoring framework for nano IoT that enables nanosensors to update a remote base station about the location and type of the detected event using only a single short pulse. Nanosensors encode different events using different center frequencies with non overlapping half power bandwidth over the entire terahertz band. Using uniform linear array (ULA) antenna, the base station localizes the events by estimating the direction of arrival of the pulse and classifies them from the center frequency estimated by spectral centroid of the received signal. Simulation results confirm that, from a distance of 1 meter, a 6th derivative Gaussian pulse consuming only 1 atto Joule can achieve localization and classification accuracies of 1.58 degree and 98.8\%, respectively.
	\end{abstract}
\vspace{-4mm}
	\section{Introduction}
\vspace{-0.2cm}
Discovery of novel nanomaterials like graphene and its derivatives has made it possible to fabricate tiny sensors measuring only a few hundred nanometers. Powered by nanostructured design, these nanosensors are capable of detecting the smallest changes in physical variables, such as pressure, vibrations, temperature, and concentrations in chemical and biological molecules. Researchers now believe that these nanosensors can transmit wireless data using graphene-based nano-antenna radiating in the terahertz band (0.1-10 THz) \cite{bA}. Such nanoscale event detection and wireless communication will open up new IoT capabilities for gathering knowledge at an unprecedented depth and scale, offering massive improvements in healthcare, agriculture, transportation, security, surveillance, industrial chemistry, and so on. Researcher are now pursuing this new direction of IoT under the banner of Internet of Nano Things (IoNT) \cite{b2} with nanoscale monitoring techniques explored for human body \cite{b3,b4}, plants \cite{b5}, chemical processes \cite{b6}, and so on. 

Graphene-based nanoantenna is a significant step forward for realizing the vision of nano IoT, but sustained event monitoring remains a major challenge due to extremely limited energy supply at nanoscale. To address the energy issue, pulse-based communication protocols are being developed for nano IoT \cite{bA} where all data is transmitted as a series of short (a few hundred femtoseconds) pulses. Use of such short pulses reduces the total energy consumption drastically compared to conventional continuous wave wireless communications, but it may not be adequate to prevent power outage at nanosensors, especially when events occur at a high rate. To conserve energy, nanosensors must be highly economical in transmitting pulses, as each pulse consumes a finite amount of energy.     

In this paper, we propose a new framework that can localize and classify events from a \textit{single pulse}. In our framework, nanosensors encode different events using different center frequencies with non overlapping half power bandwidth over the entire terahertz band. Using uniform linear array (ULA) antenna, the base station localizes the events by estimating the direction of arrival of the pulse and classifies them from the center frequency estimated by the spectral centroid of the received signal. 

The contributions in this paper can be summarized as follows:\\
	$\bullet$ We propose a unique nano IoT event monitoring framework that can localize and classify nanoscale events from a single pulse. We introduce direction of arrival (DOA) method suitable for graphene-based terahertz transceivers. We apply spectral centroid for estimating the center frequency of received pulse and classify the events.\\
	$\bullet$ We show that a single ULA operating over the entire terahertz band is not capable of localizing and classifying events when center frequencies are selected from the lower part of the band. This severely restricts the localization accuracy and the number of events that can be classified. We design a dual-ULA receiver that increases localization accuracy and classification accuracy significantly compared to a single-ULA system. \\
	$\bullet$ Using simulation, we demonstrate that, from a distance of 1 meter and using a 6th derivative Gaussian pulse consuming only 1 atto Jule, the proposed dual-ULA can localize a nanosensor within 1.58 degree and classify 5 events with 98.8\% accuracy. 
	
The rest of the paper is structured as follows. Related work is presented in section II. In section III, the terahertz channel response and molecular absorption are reviewed. We present the system model in section IV. In section V, simulations experiments are presented and discussed. Section VI concludes the paper.
\vspace{-4mm}
\section{Related Work}
Most of the energy-efficiency work for nano IoT and nano sensor networks have been carried out in the context of medium access control or routing protocols \cite{rl1,rl2}. Research on energy efficient localization and event classification for nano IoT is rare. Recently, Zarepour et al., \cite{b7} attempted single-pulse nanoscale event monitoring by harvesting the event energy and using that energy to power the pulse transmission. This allowed the base station to classify events based on the received energy of the pulse assuming that different types of events would emit different amounts of energy. This assumption holds for certain applications such as chemical reaction monitoring where different types of chemical reactions are known to emit different amounts of heat energy, which can be harvested using nanoscale thermal energy harvesters. However, Zarepour's proposal has two major limitations: (1) the base station cannot localize the event because it cannot determine which nanosensor transmitted the pulse in question, and (2) the base station cannot classify the events if all events emit similar energies. 

Later, Hassan et al., \cite{b8} addressed the first limitation of the single-pulse event monitoring framework by forcing different nanosensors to use different pulse widths, which changes the peak power of the pulse even when using the same pulse energy (same event). By detecting the peak pulse power, the base station now can tell which nanosensor has transmitted the pulse and which event has occurred. However, this improvement only solves the event localization problem, but for accurate event classification, it still relies on different events to emit different amounts of energy.  The framework we propose in this paper is widely applicable for all types of event monitoring without imposing any restrictions on event emitting energies and inclusion of energy harvesting devices inside nanosensors. This framework, therefore, is more suitable for pervasive nano IoT deployments. 
\vspace{-2mm}
\section{Terahertz Channel}
\vspace{-0.15cm}
The chemical composition of the terahertz channel affects the propagation of the pulse in two different ways. First, the propagating pulse is attenuated due to absorption of its energy by molecules in the channel. Second, the molecular absorption noise created due to re-radiation of this absorbed energy by molecules in the channel. Radiative transfer theory is used to model these two effects and is reviewed in the following subsection\cite{bA}.  
\vspace{-2mm}
\subsection{Terahertz Channel Impulse Response}
The terahertz channel response $H\left( f,d_{r}\right)$ accounts for both spreading loss $H_{spread}\left(f,d_{r} \right) $ and molecular absorption loss $H_{abs}\left(f,d_{r} \right)$ and is represented in frequency domain as 
\begin{equation}\label{eq:chresp}
H\left( f,d_{r}\right) = H_{spread}\left( f,d_{r}\right) H_{abs}\left( f,d_{r}\right)  
\end{equation}
\begin{equation}
\resizebox{0.75\columnwidth}{!}{$H_{spread}\left( f,d_{r}\right) =\left(  \frac{c_{o}}{4\pi d_{r} f_{c}}\right) \exp\left( -\frac{j 2\pi f d_{r}}{c_{o}}\right)$} 
\end{equation}
\begin{equation}
H_{abs}\left( f,d_{r}\right) = \exp(-0.5 k\left(f \right)d_{r} ) 
\end{equation}
where $f$ denotes frequency, $c_{o}$ is the velocity of light in vacuum, $f_{c}$ is the center frequency of the graphene antenna, $d_{r}$ is the path length, and $k\left( f\right) $ is the medium absorption coefficient. The medium absorption coefficient $k\left( f\right) $ of the terahertz channel at frequency $f$ composed of $Q$ type molecules is given as
\begin{equation}
\vspace{-0.25cm}
\resizebox{0.4\columnwidth}{!}{$k\left( f\right)  = \sum\limits_{q=1}^{Q}x_{q}M_{q}\left( f\right).$}
\end{equation} 
where $x_{q}$ is the mole fraction of molecule type $q$ and $M_{q}$ is the absorption coefficient of individual molecular species.
\vspace{-2mm}
\subsection{Molecular Absorption Noise}
\vspace{-0.18cm}
The ambient noise in terahertz channel is the molecular absorption noise as the thermal noise is negligible  for transceivers based on graphene\cite{bA}. The total molecular absorption noise power spectral density (p.s.d.) $S_{N}\left( f,d_{r}\right) $ affecting the transmission of pulse is the sum of background atmospheric noise p.s.d \(S_{N_{B}}\left( f,d_{r}\right)\) and the self induced noise p.s.d. $S_{N_{P}}\left(f,d_{r} \right)$ and is given as
\vspace{-2mm}
\begin{equation}
\resizebox{0.27\textwidth}{!}{$S_{N}\left( f,d_{r}\right)  = S_{N_{B}}\left(f,d_{r} \right)+S_{N_{P}}\left(f,d_{r} \right) $} 
\vspace{-1mm}
\end{equation}
\begin{equation}\label{eqn:mnm1}
\resizebox{0.43\textwidth}{!}{$S_{N_{B}}(f, d_{r}) = \lim\limits_{d_{r} \rightarrow \infty} k_{B} T_{0}\left( 1-\exp\left( -k\left(f \right)d_{r} \right) \right) \left( \frac{c_{0}}{\sqrt{4\pi}f_{c}}\right)^{2}$} 
\vspace{-1mm}
\end{equation}
\begin{equation}\label{eqn:mnm2}
\resizebox{0.43\textwidth}{!}{$S_{N_{P}}\left(f,d_{r} \right) = S_{P}\left( f\right)\left( 1-\exp\left( -k\left(f \right)d_{r} \right) \right) \left( \frac{c_{0}}{4\pi d_{r} f_{c}}\right)^{2} $} 
\vspace{-1mm}
\end{equation}
where $k_{B}$ is the Boltzmann constant, $T_{0}$ is the room temperature and $S_{P}\left( f\right) $ represents p.s.d. of transmitted pulse. 
\vspace{-1mm}
\section{System Model}
\subsection{Pulse representation for Terahertz Band Communication}
\vspace{-0.1cm}
The symbols or events at the nanoscale is represented by higher time derivative Gaussian pulse of few hundred femtoseconds duration and with a power of few \si{\micro}W. The Fourier representation of Gaussian pulse with an order of derivative $n$ is also Gaussian shaped and is represented as\cite{bA}
\vspace{-1mm}
\begin{equation}
P_{n}\left( f\right)  = a_{n}\left(j 2\pi f\right)^{n}  e^{-0.5\left( 2\pi\sigma f\right)^{2} }
\vspace{-1mm}
\end{equation}
where $a_{n}$ is the normalizing constant to adjust the pulse energy, and $\sigma$ is the standard deviation of the Gaussian pulse in seconds. Pulse duration $T_{p}$ is defined as the time interval which contains 99.99\% of pulse energy and its value is approximately equal to $10\sigma$\cite{b12}. The center frequency of Gaussian pulse increases with time derivative order and is represented as \cite{b12}
\begin{equation}
f_{c} = \frac{\sqrt{n}}{2\pi \sigma}
\end{equation} 
\subsection{Proposed Single Pulse Based Node Localization and Spectral Centroid based Decoder for Event Identification }
\begin{figure}[t]
	\centering
	\includegraphics[width=\columnwidth, height = 2.8cm]{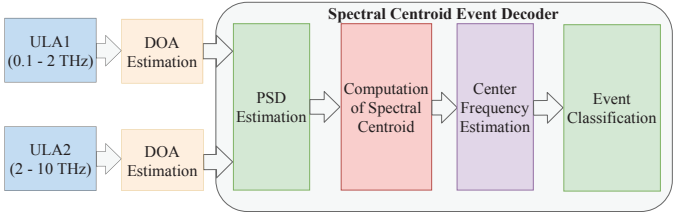}
	\vspace{-6mm}
	\caption{Proposed Dual ULA system with Spectral Centroid Decoder.}
	\label{fig:DOA_E}
	\vspace{-6mm}
\end{figure}

\vspace{-0.5mm}
The proposed system model for event localization and classification is shown in Fig. \ref{fig:DOA_E}. 
The events are localized and classified by estimating its DOA and center frequency respectively. The estimation of DOA using wideband multiple signal classification (MUSIC) algorithm is attempted in \cite{bn} and classification of center frequency using spectral centroid in the millimeter wave frequency band (100 GHz - 375 GHz) is addressed in \cite{bmb}. Using a pulse of order $n$, different events are encoded with different center frequencies that have non overlapping half power bandwidth. The center frequency of a Gaussian pulse of order $n$ is varied by changing its standard deviation. It should be noted here that, the available number of center frequencies with non overlapping bandwidth increases with increase in the order of Gaussian pulse. The steps for localizing and identifying events sensed by the nanosensor device using a single transmitted pulse is described as follows: 
	$\bullet$ The nanosensor node is localized by estimating its DOA by observing a single transmitted pulse.\\
	$\bullet$ From the estimated DOA, the p.s.d. of the transmitted pulse is estimated. \\
	$\bullet$ The spectral centroid is estimated from the estimated p.s.d. and is considered as an estimate of the center frequency of the transmitted Gaussian pulse.\\
	$\bullet$ A particular event is identified (classified) based on the estimated center frequency.

The DOA of events is estimated by employing a uniform linear array (ULA) of antennas. Suppose, if a single ULA is used to localize nanosensor devices over entire terahertz bandwidth, then nanosensor devices transmitting  Gaussian pulses with their center frequencies less than 2 THz will suffer from poor DOA estimation accuracy. The reason for this outcome is due to the decreased aperture of ULA for center frequencies below 2 THz \cite{b12_UW}. Hence, in this paper, we propose a dual ULA system as shown in Fig.\ref{fig:DOA_E}. In dual ULA system, ULA1 localizes nanosensor devices transmitting Gaussian pulses with center frequencies below 2 THz whereas ULA2 performs localization from 2 THz to 10 THz. Here, both ULA1 and ULA2 uses $N$ number of antenna elements.
\begin{figure*}[t]
	\normalsize
	\setcounter{mytempeqncnt}{\value{equation}}
	\setcounter{equation}{15}
	\begin{IEEEeqnarray}{lll}\label{eq:YYHEXP}
		\resizebox{0.9\textwidth}{0.03\columnwidth}{$\boldsymbol{Y}\left(f_{b},d_{r} \right)\boldsymbol{Y}\left(f_{b},d_{r} \right)^{H}   =  
		H\left( f_{b},d_{r}\right) \boldsymbol{a}\left( f_{b},\theta\right) \boldsymbol{P}_{n}\left( f_{b}\right)\cdot  \left(  H\left( f_{b},d_{r}\right) \boldsymbol{a}\left( f_{b},\theta\right) \boldsymbol{P}_{n}\left( f_{b}\right)\right)^{H}+\boldsymbol{V}\left( f_{b},d_{r}\right)\boldsymbol{V}^{H}\left( f_{b},d_{r}\right) +$} \\  \nonumber
		\hspace{3.6cm} \resizebox{0.7\textwidth}{0.03\columnwidth}{$H\left( f_{b},d_{r}\right) \boldsymbol{a}\left( f_{b},\theta\right) \boldsymbol{P}_{n}\left( f_{b}\right)\cdot  \left( \boldsymbol{V}\left(f_{b},d_{r} \right)  \right)^{H}+ \boldsymbol{V}\left(f_{b},d_{r}\right)  \left(  H\left( f_{b},d_{r}\right) \boldsymbol{a}\left( f_{b},\theta\right) \boldsymbol{P}_{n}\left( f_{b}\right)\right)^{H}$}\nonumber
		\vspace{-1.5mm}
	\end{IEEEeqnarray}
	\begin{equation}\label{eq:CORSIGNOI}
	\resizebox{0.9\textwidth}{!}{$\mathbb{E}\left[ \left(  H\left( f_{b},d_{r}\right) \boldsymbol{a}\left( f_{b},\theta\right) \boldsymbol{P}_{n}\left( f_{b}\right)\right)\cdot  \left( \boldsymbol{V}\left(f_{b},d_{r} \right)  \right)^{H}\right] \\
		= \left| \left|\boldsymbol{P}_{n} \right|\right|  _{2}\left( \frac{c_{0}}{4\pi d_{r} f_{c}}\right) ^{2}\frac{\sqrt{\left(1-\exp\left(-x  \right)  \right) }}{\exp\left(0.5x\right)}  \exp\left(-j\frac{2\pi f d_{r}}{c_{o}} \right) \boldsymbol{a}_{m}\left( f_{b}\right)  \boldsymbol{1}_{1\times N}$} 
	\vspace{-1mm}
	\end{equation}
	\begin{equation}\label{eq:stdeqn}
	\resizebox{0.7\textwidth}{0.03\columnwidth}{$\boldsymbol{R_{Y}}\left(f_{b},d_{r} \right)   =	\left|P_{n}\left( f_{b}\right) \right|^{2} \left|H\left( f_{b},d_{r}\right) \right|^{2} \boldsymbol{a}\left( f_{b},\theta\right)\boldsymbol{a}\left( f_{b},\theta\right)^{H} +\sigma^{2}\left( f_{b},d_{r}\right)  \boldsymbol{I}_{N}$}
	\vspace{-4mm}
	\end{equation}
	\setcounter{equation}{\value{mytempeqncnt}}
	\hrulefill
	\vspace{-6mm}
\end{figure*}
\vspace{-2mm}
\subsection{Event Localization}
\vspace{-0.5mm}
This section describes the frequency domain snapshot model and incoherent MUSIC (IMUSIC) algorithm used for estimating the DOA of a single nanosensor device. \\
$\bullet$ {\textit{Frequency Domain Snapshot Data Model}}

Without loss of generality, the frequency domain snapshot model is described for a ULA with $N$ antenna elements. In the ULA, inter-element spacing $d_{s}\:\text{m }$ is considered as half the wavelength of the highest frequency, in order to avoid spatial aliasing.   Here, it is assumed that a single nanosensor device is present in the far-field region of ULA. The path length between ULA and nanosensor device is represented as $d_{r}$. The wideband higher order Gaussian pulse received at the output of $i$th antenna element in ULA is represented as \cite{bA}. 
\vspace{-2mm} 
\begin{equation}\label{eq:datam}
y_{i}\left(t, d_{r}\right) =  p_{n}\left(t - \tau_{i} \right) * h\left(t,d_{r} \right) + v_{i}\left( t,d_{r}\right)   
\vspace{-1mm}
\end{equation}
\begin{equation}\label{eq:delay_eq}
\tau_{i} = \left(i -1\right)d_{s}\sin\left( \theta\right) /c 
\vspace{-1mm}
\end{equation}
where $h\left(t,d_{r}\right) $ is the terahertz channel impulse response between ULA and nanosensor device in $\theta$ direction. $v_{i}$ represents molecular absorption noise created between element $i$ of ULA and nanosensor device. $\tau_{i}$ represents the propagation delay between $i^{\text{th}}$ and reference antenna in ULA. For DOA estimation using a single higher order Gaussian pulse, the output across the ULA is observed for time duration $\Delta T$, which is slightly larger than its total duration $T_{p}$. For observation time $\Delta T$ longer than the propagation time of the Gaussian pulse across the ULA, the Fourier representation of  \eqref{eq:datam} is given as\cite{b13}
\vspace{-1mm}
\begin{IEEEeqnarray}{lll}
Y_{i}\left( f_{b},d_{r}\right) =  e^{-j2\pi f_{b} \tau_{i}} P_{n}\left( f_{b}\right) H\left( f_{b},d_{r}\right) + V_{i,}\left( f_{b},d_{r}\right), \\ \nonumber
\hspace{5.3cm}\text{for}\; b = 0,\cdots, L
\vspace{-1mm}
\end{IEEEeqnarray}
where $f_{b} $ is the frequency bin, $P_{n}\left( f_{b}\right) $, $H\left( f_{b},d_{r}\right) $, and $V_{i,}\left( f_{b},d_{r}\right)$ are Fourier coefficients of Gaussian pulse, terahertz channel impulse response and molecular absorption noise respectively. Further the output of array can be observed for $K$ non-overlapping time interval $\Delta T$. Here, $K$ is called as frequency snapshot number. The value of frequency snapshot $K$ is set to 1 for DOA estimation using single pulse. The number of frequency bins $L$ in observation time $\Delta T$ is given as \cite{b13}
\vspace{-1mm}
\begin{equation}\label{eq:timbw}
L = \lfloor B \cdot \Delta T \rfloor +1
\vspace{-1mm}
\end{equation}
where $\lfloor \cdot  \rfloor$ is the floor operator, $B$ is the terahertz channel bandwidth. Now, the Fourier coefficients at frequency $f_{b}$ across $N$ sensors for $K$ number of frequency snapshots is represented in matrix form as 
\vspace{-1.5mm}
\begin{equation}\label{eq:datafm_vec}
\resizebox{0.4\textwidth}{!}{$\boldsymbol{Y}\left(f_{b},d_{r} \right) =  H\left( f_{b},d_{r}\right) \boldsymbol{a}\left(f_{b},\theta \right) \boldsymbol{P}_{n}\left(f_{b} \right) + \boldsymbol{V}\left( f_{b},d_{r}\right)$} 
\end{equation}
where $\boldsymbol{Y}\left(f_{b},d_{r} \right) \in \mathbb{C}^{N \times K}$, $\boldsymbol{V}\left( f_{b},d_{r}\right)\in \mathbb{C}^{N \times K}$ and\\ $\boldsymbol{P}_{n}\left(f_{b} \right)\overset{\Delta}{=}\left[ {P}_{n1}\left(f_{b} \right),\cdots,{P}_{nK}\left(f_{b} \right)\right] $.\\$\boldsymbol{a}\left(f_{b},\theta \right) =\left[1, e^{-j2\pi f_{b} \tau_{1}}, \cdots, e^{-j2\pi f_{b} \tau_{N}}\right]^{T} $ is the array manifold vector. The covariance matrix $\boldsymbol{R_{Y}}\left(f_{b},d_{r} \right)$ of $\boldsymbol{Y}\left(f_{b},d_{r} \right)$ is given as 
\vspace{-2mm}
\begin{equation}\label{eq:covmat}
\resizebox{0.6\columnwidth}{!}{$\boldsymbol{R_{Y}}\left(f_{b},d_{r} \right)  = \mathbb{E}\left[ \boldsymbol{Y}\left(f_{b},d_{r} \right)\boldsymbol{Y}\left(f_{b},d_{r} \right)^{H}\right]$}
\vspace{-2mm} 
\end{equation}
where $\left(\cdot \right) ^{H}$ denotes conjugate transpose and $\mathbb{E}\left[ \cdot\right] $ represents expectation. The term \resizebox{0.35\columnwidth}{!}{$\left[ \boldsymbol{Y}\left(f_{b},d_{r} \right)\boldsymbol{Y}\left(f_{b},d_{r} \right)^{H}\right] $} in \eqref{eq:covmat} is represented in \eqref{eq:YYHEXP}. Taking expectation on \eqref{eq:YYHEXP} and using \eqref{eq:chresp} and \eqref{eqn:mnm2}, the expectation of third term on right hand side of \eqref{eq:YYHEXP} simplifies to \eqref{eq:CORSIGNOI}. Here it is assumed that background atmospheric noise has zero mean. In \eqref{eq:CORSIGNOI}, $\left| \left|\cdot \right| \right|_{2} $ denotes $l_{2}$ norm,  $x = k\left( f_{b}\right)\cdot d_{r}$  and $\boldsymbol{1}_{1\times N}$ is ones vector of size $1\times N$. The term $\left(\frac{c_{o}}{4\pi d_{r} f_{c}} \right)^{2} \ll 1 $, as center frequency $f_{c}$ of the Gaussian pulse is in terahertz range. Further, since the molecular noise temperature is low for smaller path lengths $d_{r}$ \cite{bA}
\setcounter{mytempeqncnt}{\value{equation}}
\setcounter{equation}{18}
\vspace{-1mm}
\begin{equation}
\frac{\sqrt{\left(1-\exp\left(-x  \right)  \right) }}{\exp\left(0.5x\right)} \approx 0
\vspace{-1mm}
\end{equation}
Based on these assumptions \eqref{eq:CORSIGNOI} is approximated as zero. Similar arguments can be made for the fourth term in \eqref{eq:YYHEXP} and can approximated to zero.
Based on the above assumptions and for $K = 1$, \eqref{eq:covmat} is simplified as \eqref{eq:stdeqn}. In \eqref{eq:stdeqn},  $\boldsymbol{I}_{N}$ is the identity matrix of size $N\times N$ and  \resizebox{0.5\columnwidth}{!}{$\mathbb{E}\left[ \boldsymbol{V}\left( f_{b},d_{r}\right)\boldsymbol{V}^{H}\left( f_{b},d_{r}\right) \right] =\sigma^{2}\left( f_{b},d_{r}\right)$} is the noise variance around narrow frequency sub-band centered at frequency $f_{b}$. Eqn. \eqref{eq:stdeqn} is same as the covariance matrix at the output of ULA assuming noise to be independent of Gaussian pulses emitted by nanosensor devices. $\sigma^{2}\left( f_{b},d_{r}\right)$ is computed as 
\vspace{-2mm}
\begin{equation}
\resizebox{0.45\columnwidth}{!}{$\sigma^{2}\left( f_{b},d_{r}\right) = \int S_{N}(f_{b},d_{r}) df$}
\vspace{-2mm}
\end{equation}
$\bullet$ \textit{DOA estimation of Gaussian Pulses}

The IMUSIC algorithm can perform DOA estimation even with a single pulse is due to low molecular absorption noise for path lengths below 0.5 m \cite{b13_T}. The IMUSIC wideband DOA estimation technique is given as \cite{b13_IM}
\vspace{-1.5mm}
\begin{equation}\label{eq:IMUS}
\resizebox{0.43\textwidth}{!}{$P_{\text{IMUSIC}}( \hat{\theta}, d_{r})  = \sum\limits_{b=1}\limits^{L}\frac{\boldsymbol{a}^{H}\left(f_{b},\theta \right)\boldsymbol{a}\left(f_{b},\theta \right) }{\boldsymbol{a}^{H}\left(f_{b},\theta \right)\boldsymbol{E}_{n}\left(f_{b},d_{r} \right)\boldsymbol{E}_{n}^{H}\left(f_{b},d_{r} \right)\boldsymbol{a}\left(f_{b},\theta \right)} $}
\vspace{-1mm}
\end{equation}	
where $\boldsymbol{E}_{n}\left(f_{b},d_{r} \right)$  is the  noise eigenvector matrix which is obtained from eigenvalue decomposition of $\boldsymbol{R_{Y}}\left(f_{b},d_{r} \right) $.  Eqn. \eqref{eq:IMUS} is called as IMUSIC spectrum and it is observed that, the quality of DOA estimate depends on communication distance between nanosensor device and ULA. The DOA estimate from IMUSIC spectrum is estimated as 
\vspace{-2.25mm}
\begin{equation}\label{eq:tht_est}
\resizebox{0.55\columnwidth}{!}{$\hat{\theta} \left( d_{r}\right) = \argmax\limits_{\theta}\left[ P_{\text{IMUSIC}}( \hat{\theta}, d_{r})\right] $}
\vspace{-1mm}
\end{equation}
\vspace{-1mm}
Further, the received covariance matrix at each frequency bin $f_{b}$ is estimated as
\vspace{-1mm}
\begin{equation}
\resizebox{0.6\columnwidth}{!}{$\boldsymbol{\hat{R}_{Y}}\left(f_{b},d_{r} \right) = \frac{1}{K}\boldsymbol{Y}\left(f_{b},d_{r} \right) \boldsymbol{Y}^{H}\left(f_{b},d_{r} \right)$}
\vspace{-1mm}
\end{equation}
\subsection{Power Spectral Density and central frequency estimation}
\begin{figure}
	\centering
	\includegraphics[width=\columnwidth, height = 3.25cm]{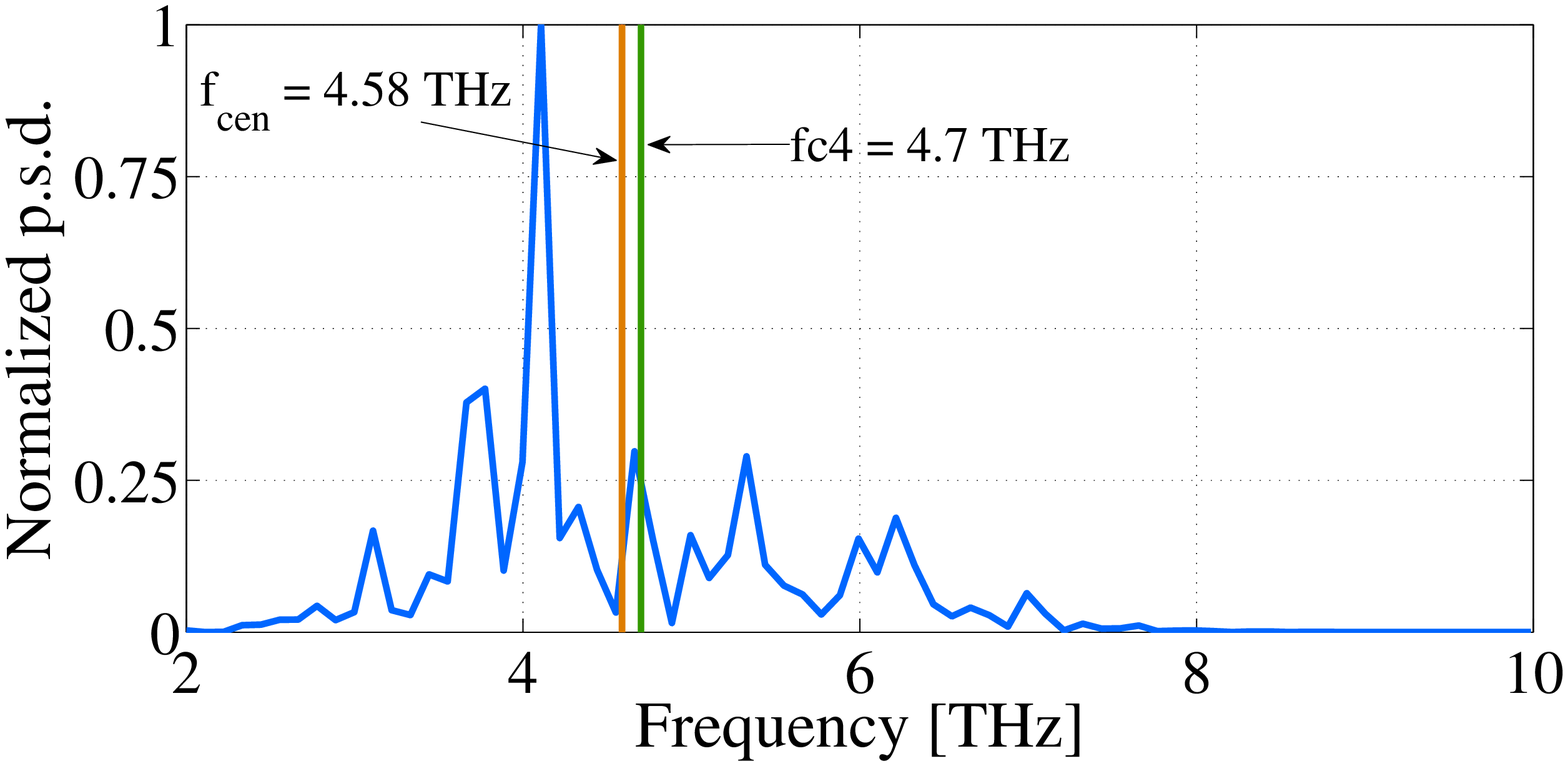}
	\vspace{-2mm}
	\caption{Estimated p.s.d. (blue line) for 1 aJ Gaussian pulse with center frequency 4.7 THz  transmitted from a distance of 5 mm. While the center frequency estimated by the peak of blue line is far from the actual value, the proposed spectral centroid estimates the center frequency ($f_{cen}$) at 4.58 THz.}
	\label{fig:fcent}
	\vspace{-6mm}
\end{figure}

\vspace{-0.5mm}
This section describes the estimation of p.s.d. and center frequency of the higher order Gaussian pulse transmitted by the nanosensor device. Since the molecular absorption noise is negligible for small path lengths, the noise term in \eqref{eq:stdeqn} can be neglected and the estimated p.s.d. is obtained as 
\vspace{-1mm}
\begin{equation}
\resizebox{0.6\columnwidth}{!}{$\hat{S}_{n}\left( f_{b}\right)  =\left(  \hat{\boldsymbol{a}}\left(f_{b},\theta \right) \right) ^{\dagger}\boldsymbol{\hat{R}_{Y}}\left(f_{b},d_{r} \right)\left(  \hat{\boldsymbol{a}}\left(f_{b},\theta \right) ^{H}\right) ^{\dagger}$}
\vspace{-1mm}
\end{equation}
where $\left( \cdot\right) ^{\dagger}$ represents pseudoinverse operator and $\hat{\boldsymbol{a}}\left(f_{b} \right) $ is the array steering vector computed using DOA estimate $\hat{\theta}$. The estimated p.s.d. by ULA2 for sixth order Gaussian pulse with center frequency 4.7 THz obtained in a single simulation trial is shown in Fig. \ref{fig:fcent}. It is observed from Fig. \ref{fig:fcent} that, from the estimated p.s.d. it is difficult to locate the center frequency of the Gaussian pulse. A possible explanation for this outcome is the molecular absorption loss of propagating waves in the terahertz channel. This depends on the value of resonant peaks (see Fig.\ref{fig:ABS_COEFF}) in the molecular absorption coefficient of the channel. To overcome this problem, the center frequency is estimated by computing the spectral centroid of the estimated p.s.d. It is defined as the center of mass of the amplitude or power spectrum. In literature spectral centroid is used for speaker recognition\cite{b13_1}. The spectral centroid is defined as
\vspace{-1mm}
\begin{equation}
\resizebox{0.75\columnwidth}{!}{$f_{cen} = \frac{\sum\limits_{b = 1}\limits^{L}f_{b}\cdot\hat{S}_{n}\left( f_{b}\right)\cdot\Delta f}{\sum\limits_{b = 1}\limits^{L}\hat{S}_{n}\left( f_{b}\right)\cdot\Delta f} = \frac{\sum\limits_{b = 1}\limits^{L}f_{b}\cdot\hat{S}_{n}\left( f_{b}\right)}{\sum\limits_{b = 1}\limits^{L}\hat{S}_{n}\left( f_{b}\right)}$}
\vspace{-1mm}
\end{equation}
here $\Delta f$ represents width of frequency bin and $\hat{S}_{n}\left( f_{b}\right)\cdot\Delta f$ represents the estimated power spectrum. Based on the computed spectral centroid $f_{cen}$, the center frequency $f_{c{i}}$ representing event $i$ is estimated according to the following rule
\vspace{-1mm}
\begin{equation}\label{eq:freqest}
\hat{f_{c}} = f_{c_{i}}\: \text{if} \left | f_{cen}-f_{c_{i}} \right |\leq \left | f_{cen}-f_{c_{j}} \right |\: \forall \: j\neq i
\vspace{-1mm}
\end{equation} 
\vspace{-7mm}
\section{Simulation Results}
\vspace{-1mm}
In this section, simulation results are presented for event localization and classification for single and dual ULA. IMUSIC and spectral centroid estimation algorithm as explained in the previous section is implemented using MATLAB 2014a.
\vspace{-2mm}
\subsection{Parameters and Performance Metrics}
\begin{figure}
	\vspace{-7mm}
	\centering
	\includegraphics[width=0.8\columnwidth, height = 2.7cm]{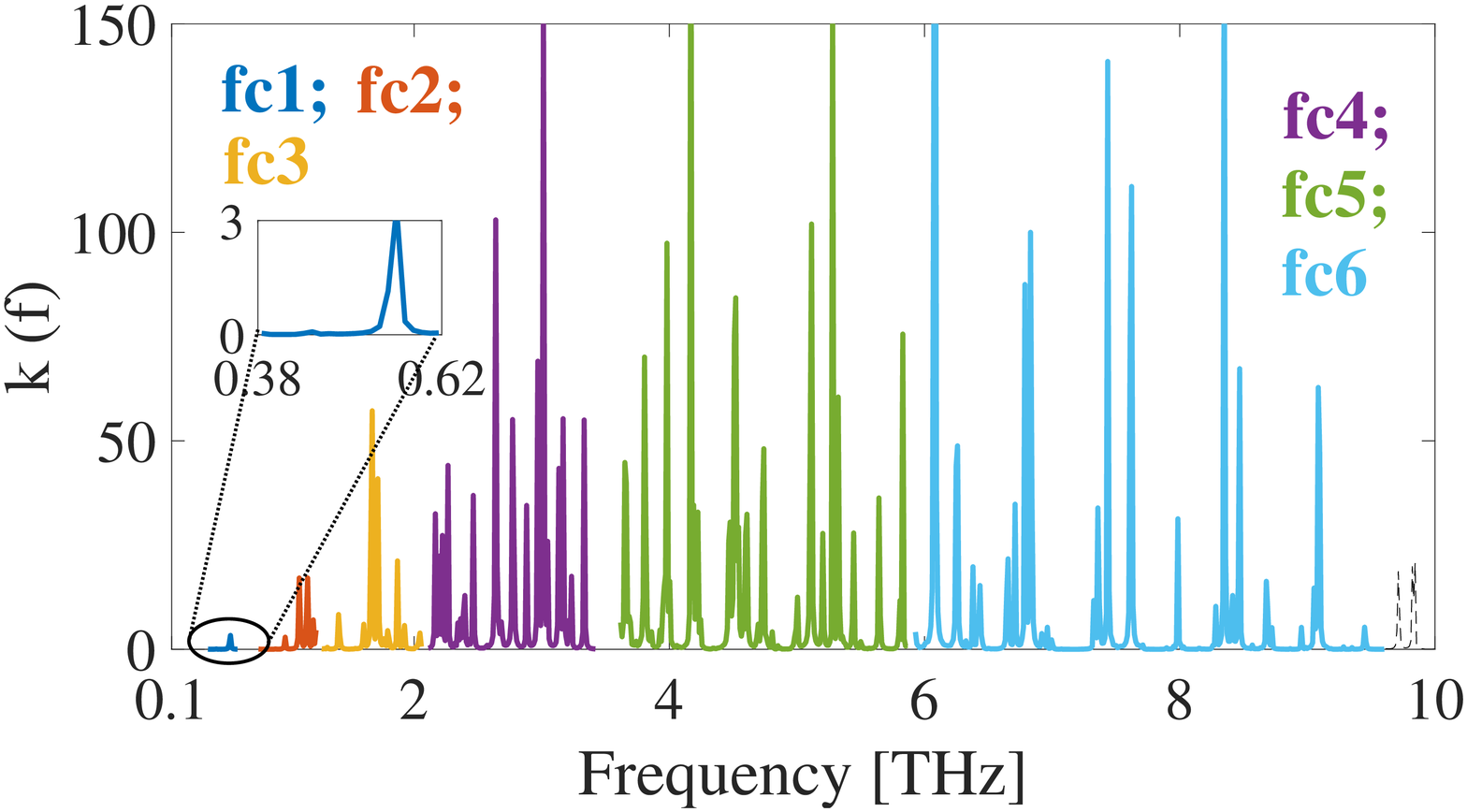}
	\vspace{-4mm}
	\caption{Molecular absorption coefficient within half power bandwidth of different center frequencies $f_{c_{i}}$ for standard air medium with 1.86\% concentration of water vapor molecules.}
	\label{fig:ABS_COEFF}
	\vspace{-4mm}
\end{figure}

\vspace{-1mm}
The terahertz channel frequency band is considered from 0.1 THz to 10 THz and the high-resolution transmission molecular absorption (HITRAN) database \cite{b14} is used to obtain the molecular absorption coefficient $k\left( f\right) $ of the terahertz channel for standard summer air with $1.86 \%$ concentration of water vapor. The specifications for single and dual ULA are shown in Table \ref{tab:TABLE1}. The transmitting nanosensor device is assumed to be located in far-field region of ULAs with DOA $-18.525^{\circ}$\footnote{Similar performance can be observed for other angles as long as they are not too close to the array axis} and transmitting sixth order Gaussian pulses at one of the six different center frequencies depending on the type of event sensed. Table \ref{tab:TABLE2} shows these six different center frequencies $\left\lbrace f_{c_{i}}\right\rbrace_{i=1}^{6} $ along with their half power frequencies and pulse duration. It is observed from Table \ref{tab:TABLE2} that, Gaussian pulse with lowest center frequency has the largest pulse duration when compared to other center frequencies.  Hence, in order to fit a single Gaussian pulse of given order with different center frequencies within the observation interval, $\Delta T$ is set to a value slightly larger than the pulse duration of the Gaussian pulse with lowest center frequencies. Thus in the simulation, the value of $\Delta T$ is selected as 9 ps. The required sampling rate at ULA is 20 THz, which is greater than twice the maximum frequency 9.61 THz, but the maximum sampling rate of current data converters is 100 GHz\cite{b15_sam1}. To overcome this limitation, sub-Nyquist sampling methods like the finite rate of innovation can be used to obtain Fourier coefficients at sampling rates less than the Nyquist rate\cite{b15_sam2}. Fig.\ref{fig:ABS_COEFF} plots the molecular absorption coefficient within different center frequencies $f_{c_{i}}$ and reveals that the molecular resonance peaks are most concentrated within the half-power bandwidth of the fourth center frequency (in blue).
\begin{table}[t]
	\centering
	\caption{Specification for single and dual ULA}
	\vspace{-0.25cm}
	\resizebox{\columnwidth}{0.065\columnwidth}{\begin{tabular}{|c|c|c|c|c|c|c|}
			\hline
			\multicolumn{2}{|c|}{}                              & $N$ & $d_{s} = \frac{\lambda_{\text{min}}}{2}$ & $f_{l}$ [THz]  & $f_{h}$ [THz] &$ L$  \\ \hline
			\multicolumn{2}{|c|}{\textbf{Single ULA}}           & 8 & $\SI{15}{\micro\meter}$    & 0.1 & 10 & 91 \\ \hline
			\multirow{2}{*}{\textbf{Dual ULA}} & \textit{ULA 1} & 8 & $\SI{75}{\micro\meter}$   & 0.2 & 2  & 19 \\ \cline{2-7} 
			& \textit{ULA 2} & 8 & $\SI{15}{\micro\meter}$   & 2   & 10 & 73 \\ \hline
	\end{tabular}}
	\label{tab:TABLE1}
	\vspace{-5mm}
\end{table}
\begin{table}[t]
	\centering
	\caption{Half power frequency of sixth order Gaussian pulse at different center frequencies}
	\vspace{-2mm}
	\resizebox{\columnwidth}{0.09\columnwidth}{\begin{tabular}{|c|c|c|c|c|}
			\hline
			$f_{c}$ [THz]   & $T_{p} = 10\sigma$ [ps]   & $f_{l}$ [THz]  & $f_{h}$ [THz]   & $B_{3dB}$ [THz]\\ \hline
			0.5  & 7.8  & 0.38 & 0.62 & 0.23 \\ \hline
			1    & 3.9  & 0.77 & 1.24 & 0.47 \\ \hline
			1.65 & 2.3  & 1.27 & 2.06 & 0.79 \\ \hline
			2.75  & 1.41 & 2.11 & 3.43 & 1.31 \\ \hline
			4.7  & 0.82 & 3.61 & 5.87 & 2.25 \\ \hline
			7.7  & 0.50 & 5.92 & 9.61 & 3.68 \\ \hline
	\end{tabular}}
	\label{tab:TABLE2}
\vspace{-7mm}
\end{table}
The estimation accuracy of the parameters, DOA $\hat{\theta}$ and center frequency $\hat{f}_{c}$ for single sensor node is measured in terms of root mean square error (RMSE) and is defined as
\vspace{-1mm}
\begin{equation}
\resizebox{0.5\columnwidth}{0.07\columnwidth}{$RMSE_{g} = \sqrt{\frac{1}{N_{run}}\sum\limits_{i=1}^{N_{run}}\left (\hat{g}\left ( i \right ) -g \right )^{2}}$}
\vspace{-2mm}
\end{equation}
where $N_{run}$ is the total number of single pulse transmissions, $\hat{g}\left ( i \right )\in \left\lbrace \hat{\theta}, \hat{f}_{c}\right\rbrace $ is the estimate of the parameters in $i^{\text{th}}$ simulation run and its corresponding true value is $g \in \left\lbrace \theta, f_{c}\right\rbrace $.  Finally, the accuracy of event classification is defined as true positive rate (TPR), which is obtained as the ratio of number of correct classifications divided by the total number of events (pulse transmissions) simulated. The total number of single pulse transmissions $N_{run}$ is set to 100.
\vspace{-2.5mm}
\begin{figure}
	\vspace{-1mm}
	\centering
	\subfigure[~]{
		\includegraphics[width=0.5\columnwidth, height = 3cm]{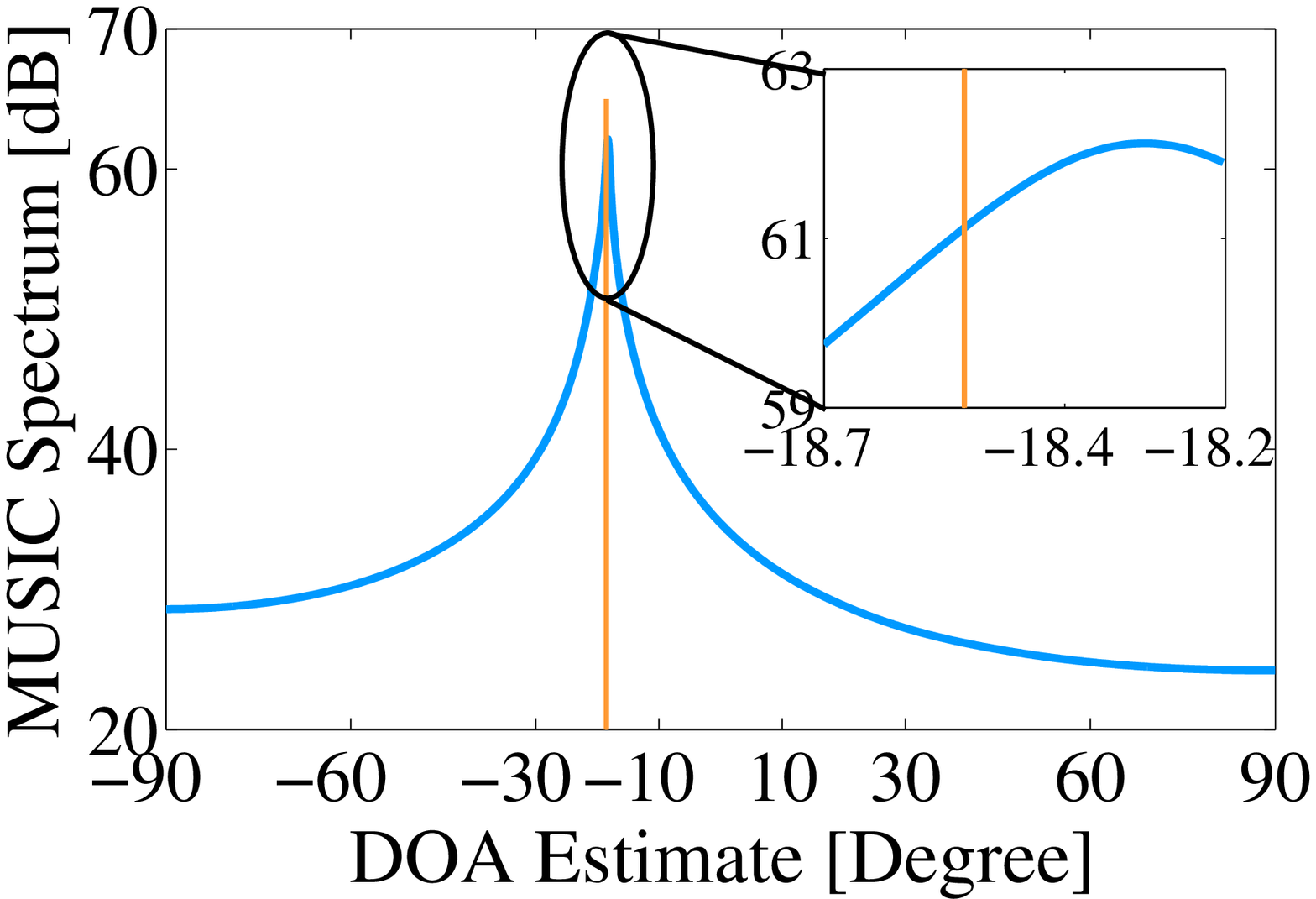}
		\label{fig:ULA_1}
	}
	\hspace{-0.8cm}
	\subfigure[~]{
		\vspace{-2mm}
		\includegraphics[width=0.5\columnwidth, height = 3cm]{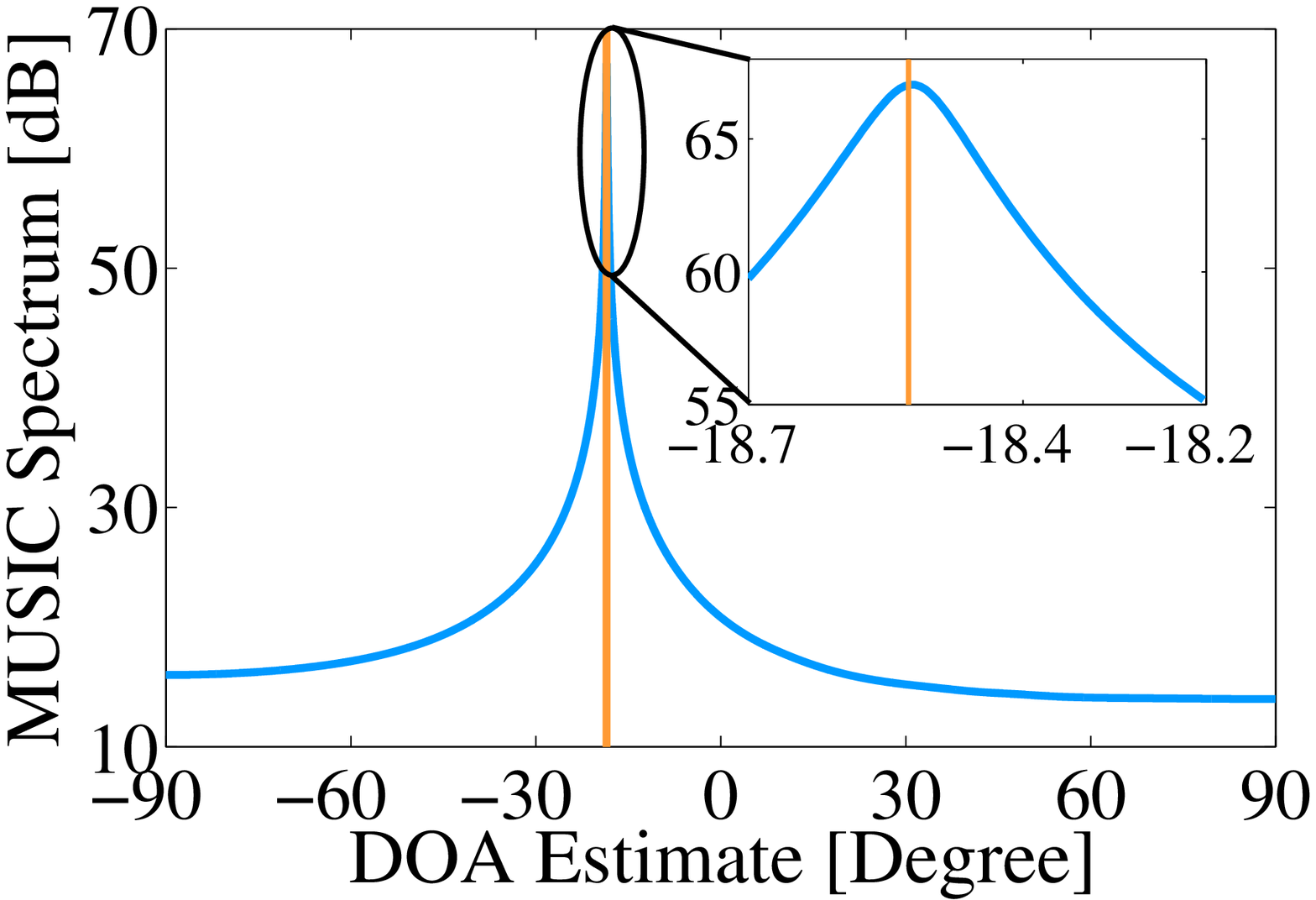}
		\label{fig:ULA_FU} }
	\vspace{-5mm}
	\caption{IMUSIC Spectrum estimated at single (a) and dual (b) ULA (ULA1) for distance 5 mm, and center frequency of 0.5 THz.}
	\vspace{-2mm}
	\label{fig:ULA_COMP}
	\vspace{-3mm}
\end{figure}
\subsection{DOA Estimation}
\vspace{-1mm}
Fig. \ref{fig:ULA_COMP} compares IMUSIC spectrum obtained from single ULA and dual ULA (ULA1). We can clearly observe that the peak is closer to the DOA value $\left( -18.525^{\circ}\right) $ for dual ULA compared to that of single ULA. A possible explanation for poor DOA estimation by single ULA is because inter-element spacing in single ULA is too small (15 $\mu$m compared to 75 $\mu$m in ULA1) for center frequencies below 2 THz. 

\begin{figure}[ht]
	\centering
	\includegraphics[width=0.75\columnwidth, height = 2.5cm]{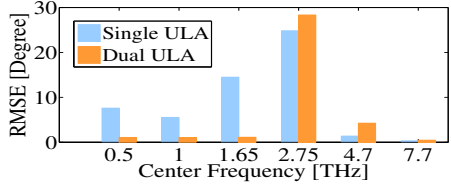}
	\vspace{-2mm}
	\caption{Center Frequency versus RMSE for DOA for path length 1 m}
	\label{fig:BAR_GRA}
	\vspace{-5mm}
\end{figure}

\begin{figure}[ht]
	\centering
	\subfigure[~]{
		\includegraphics[width=0.48\columnwidth, height = 2.3cm]{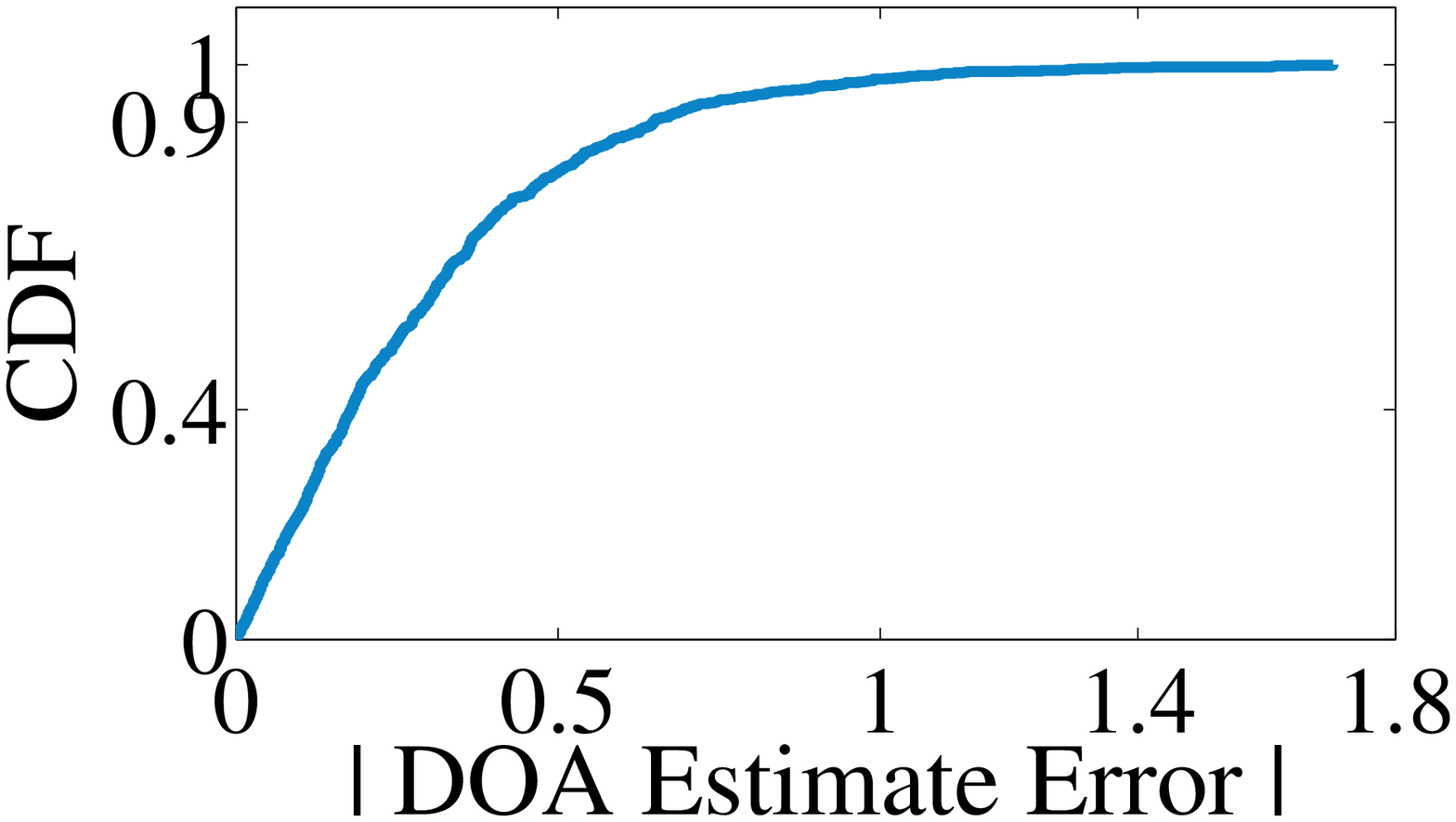}
		\label{fig:CDF}
	}
	\hspace{-0.7cm}
	\subfigure[~]{
		\includegraphics[width=0.48\columnwidth, height = 2.3cm]{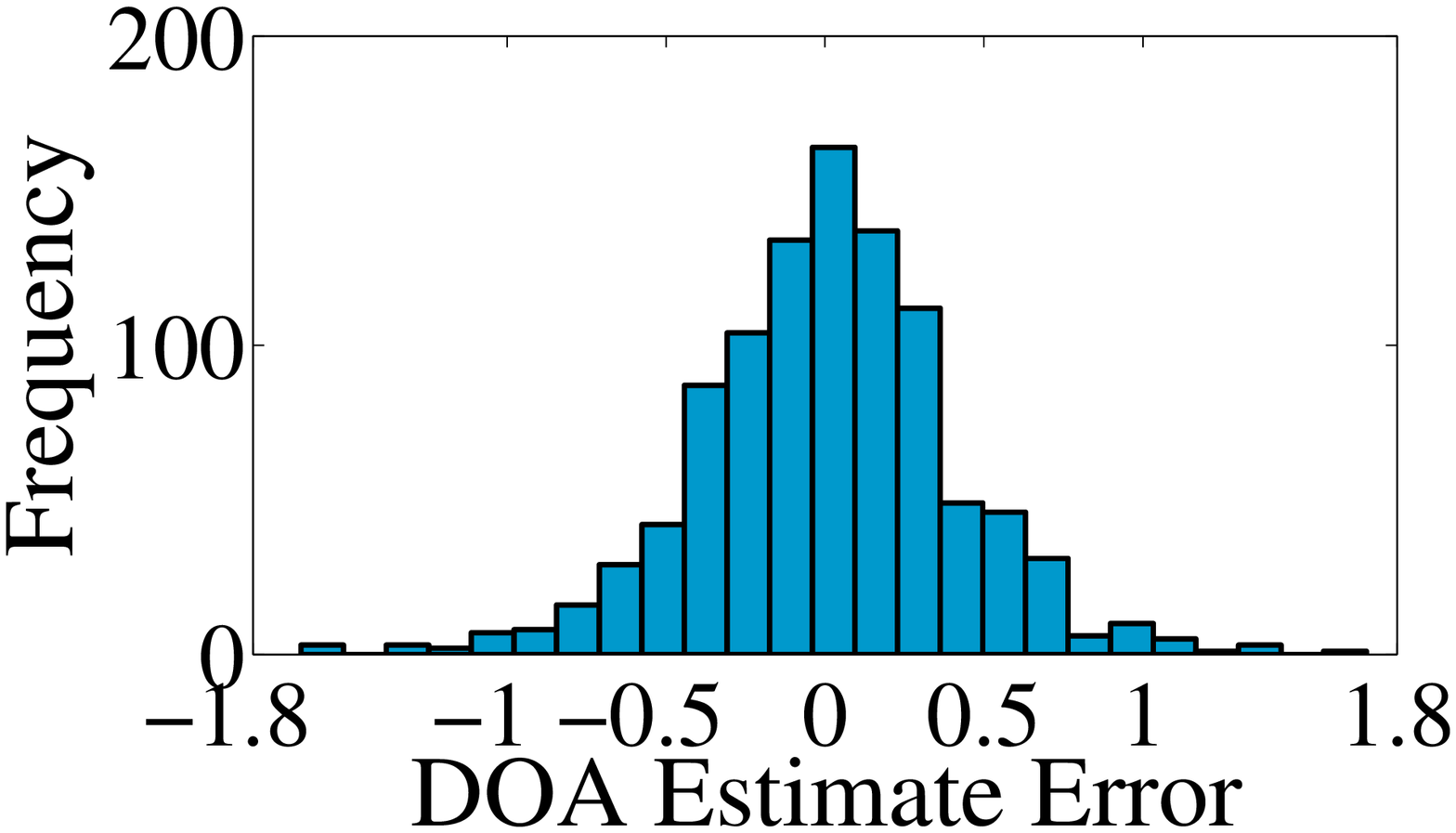}
		\label{fig:PDF} }
	\vspace{-3mm}
	\caption{CDF (a) and histogram (b) for sixth order Gaussian pulse using ULA2 in dual ULA for path length of 1 m and center frequency of 7.7 THz.}
	\label{fig:PDF_CDF}
	\vspace{-6.5mm}
\end{figure}

Figs. \ref{fig:RMTF05} and \ref{fig:RMTF1} shows the DOA estimation performance as a function of path length $d_{r}$. We observe that for lower frequencies (Figs. \ref{fig:RMTF05}), dual ULA provides much better localization accuracy at all distances. In contrast, for higher frequencies (Fig. \ref{fig:RMTF1}), the differences between single and dual ULA diminishes, but the differences between different frequencies become more prominent. Fig. \ref{fig:BAR_GRA} captures DOA performance at a distance of 1m. The 4th frequency performs very bad due to the high density of resonance peaks (see Fig. \ref{fig:ABS_COEFF}) in its half power bandwidth. By excluding $f_{c_{4}}$, dual ULA provides an average RMSE of $1.58^{\circ}$ across all center frequencies while single ULA  achieves an average RMSE of $5.86^{\circ}$. Finally, Fig. \ref{fig:PDF_CDF} confirms that for dual ULA, DOA estimate is close to the true DOA of nanosensor most of the time and the absolute DOA error is less than $1^{\circ}$ for 97\% of the time.
\vspace{-2mm}
\subsection{Event Classification}
\begin{figure*}[ht]
	\centering
	\subfigure[~]{
		\includegraphics[width=0.54\columnwidth, height = 4cm]{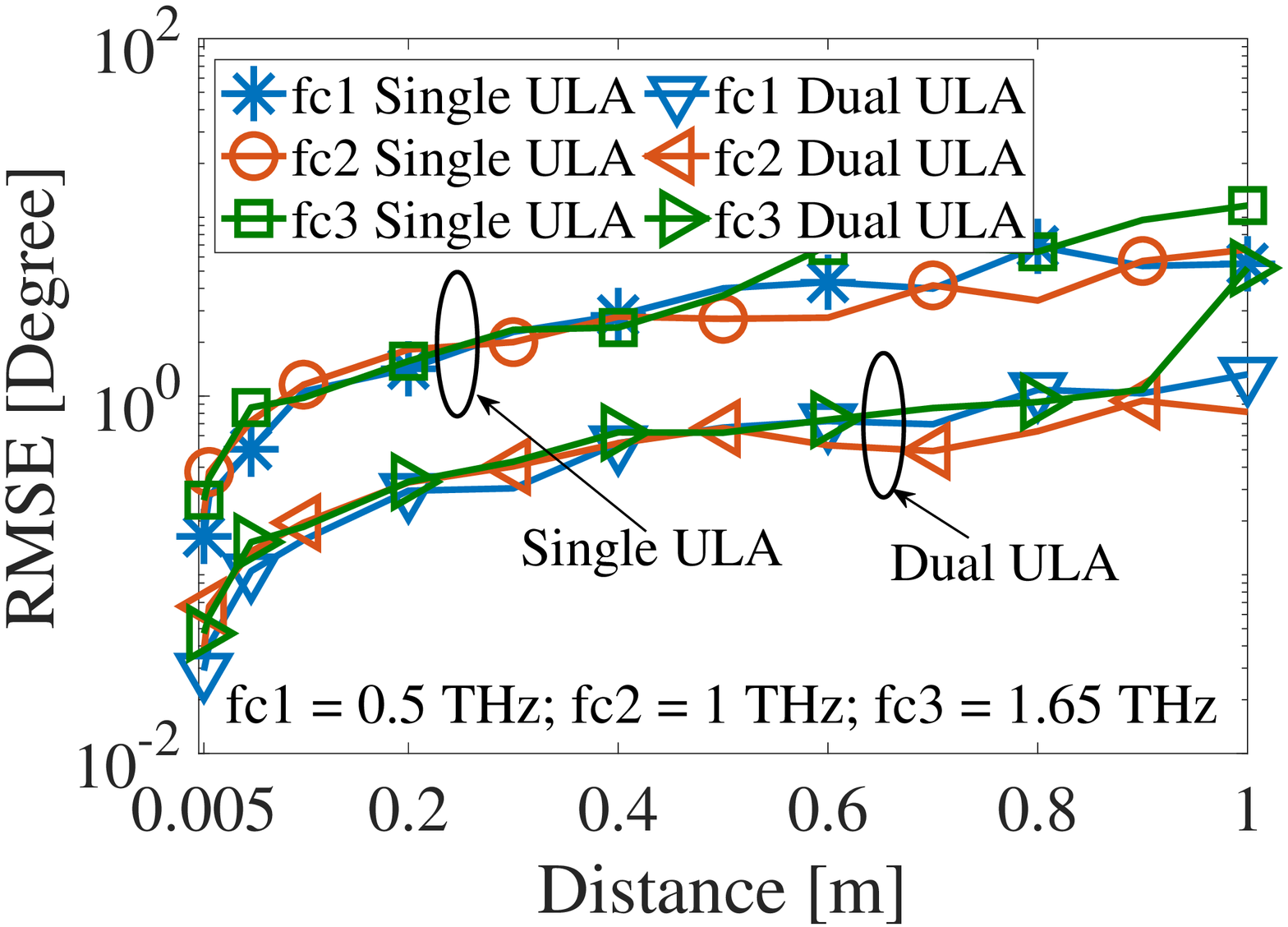}
		\label{fig:RMTF05}
	}
	\hspace{-8.5mm}
	\subfigure[~]{
		\includegraphics[width=0.54\columnwidth, height = 4cm]{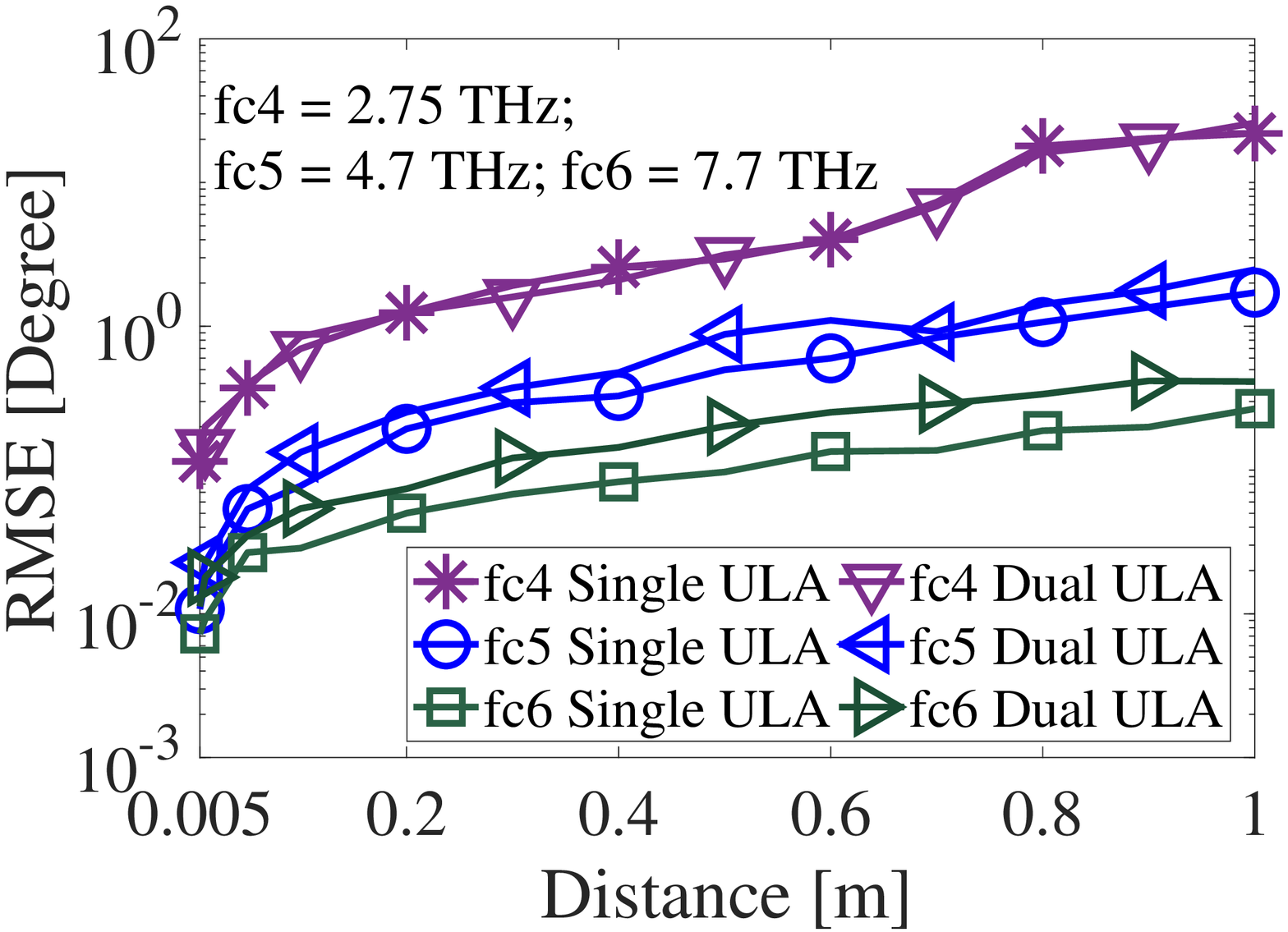}
		\label{fig:RMTF1}
	}
	\hspace{-8.5mm}
	\subfigure[~]{
		\includegraphics[width=0.54\columnwidth, height = 4cm]{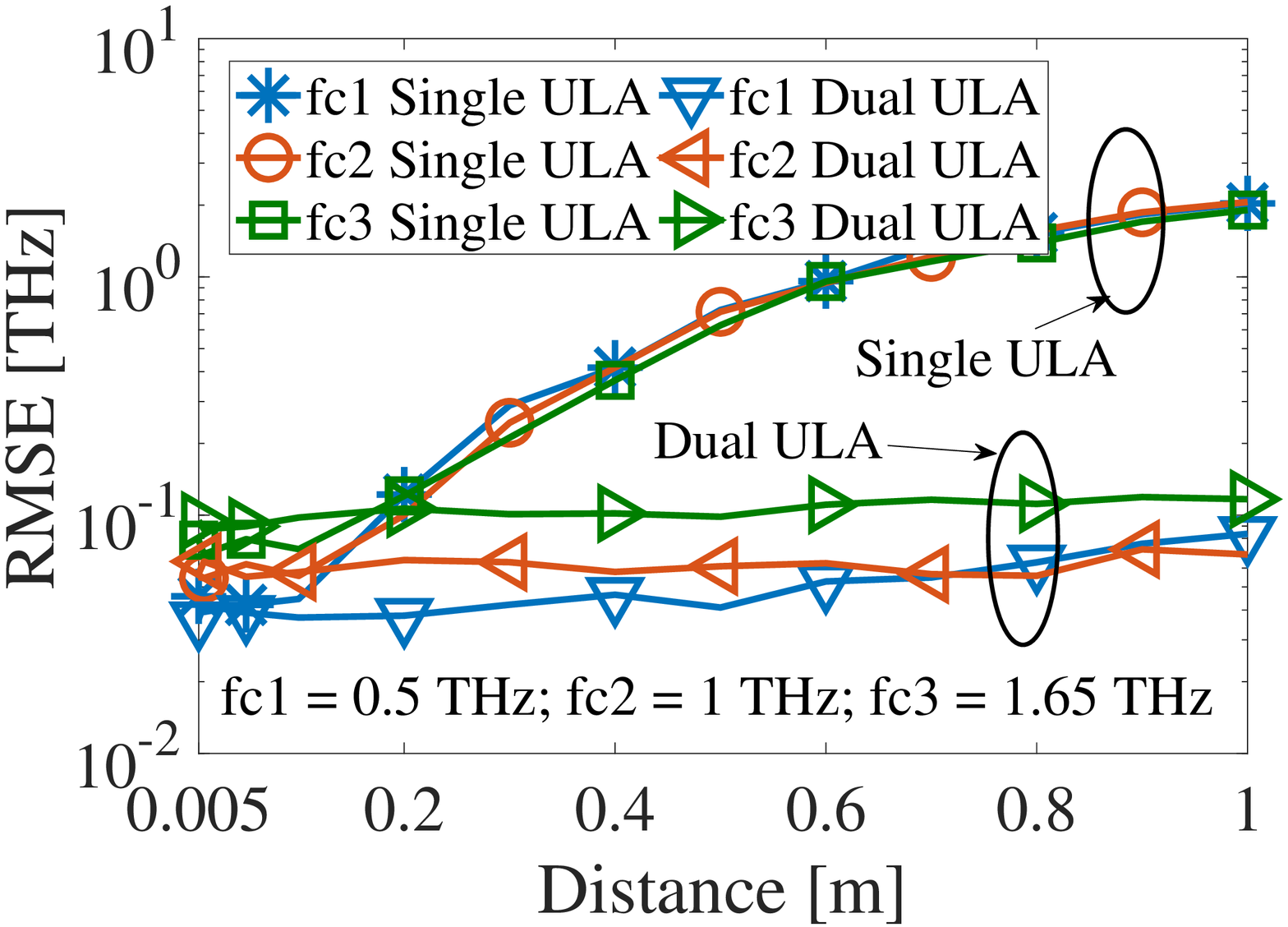}
		\label{fig:RMFRF05}
	}
	\hspace{-8.5mm}
	\subfigure[~]{
		\includegraphics[width=0.54\columnwidth, height = 4cm]{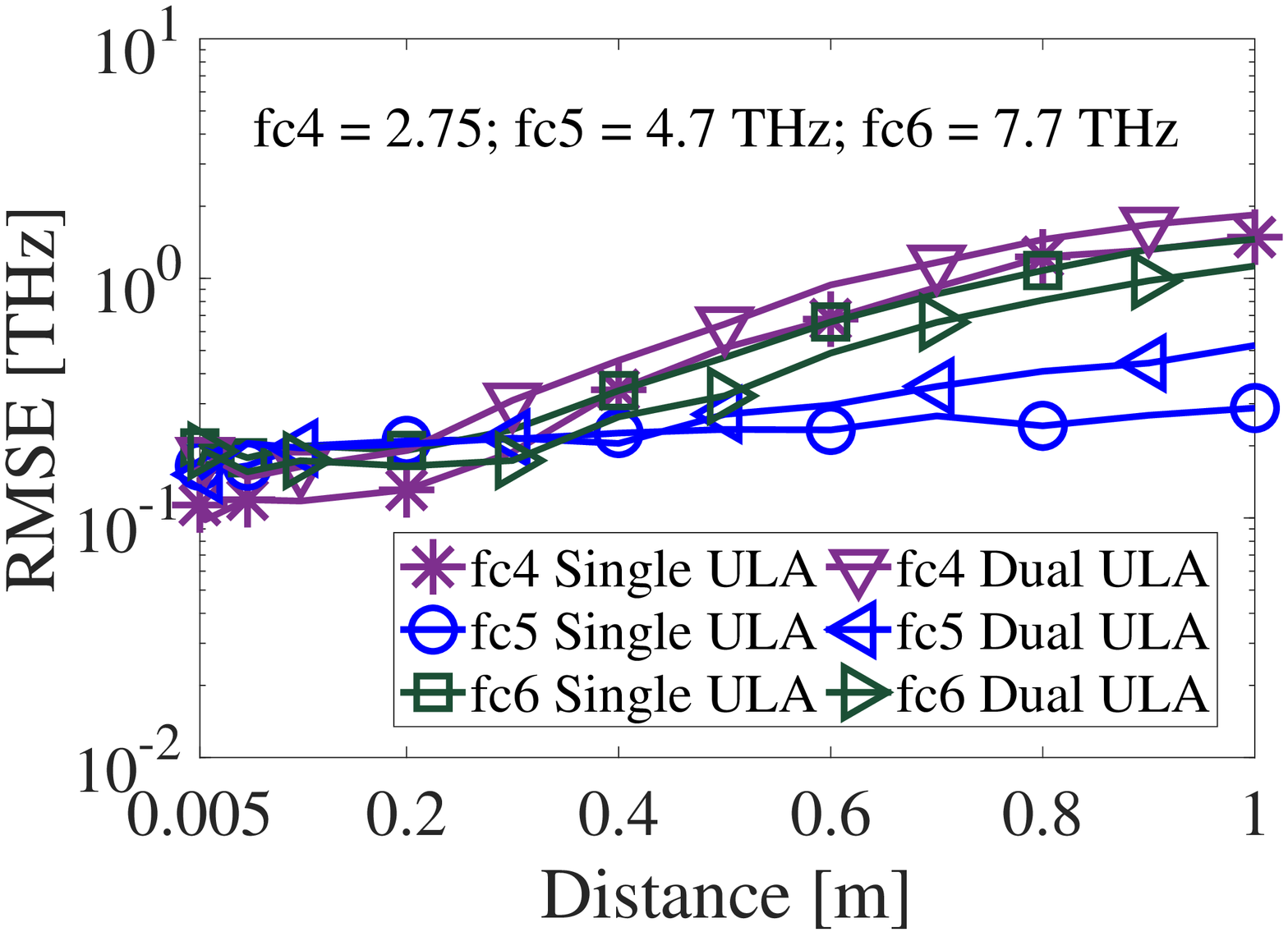}
		\label{fig:RMFRF1}
	}
	\vspace{-2mm}
	\caption{(a) and (b) Comparison of DOA, and (c) and (d) Center frequency estimation accuracy for Single and Dual ULA.
		\label{fig:SigDuRM}}
	\vspace{-8mm}
\end{figure*}

\begin{figure}[ht]
	\centering
	\subfigure{
		\includegraphics[width=0.34\columnwidth,height=2.5cm]{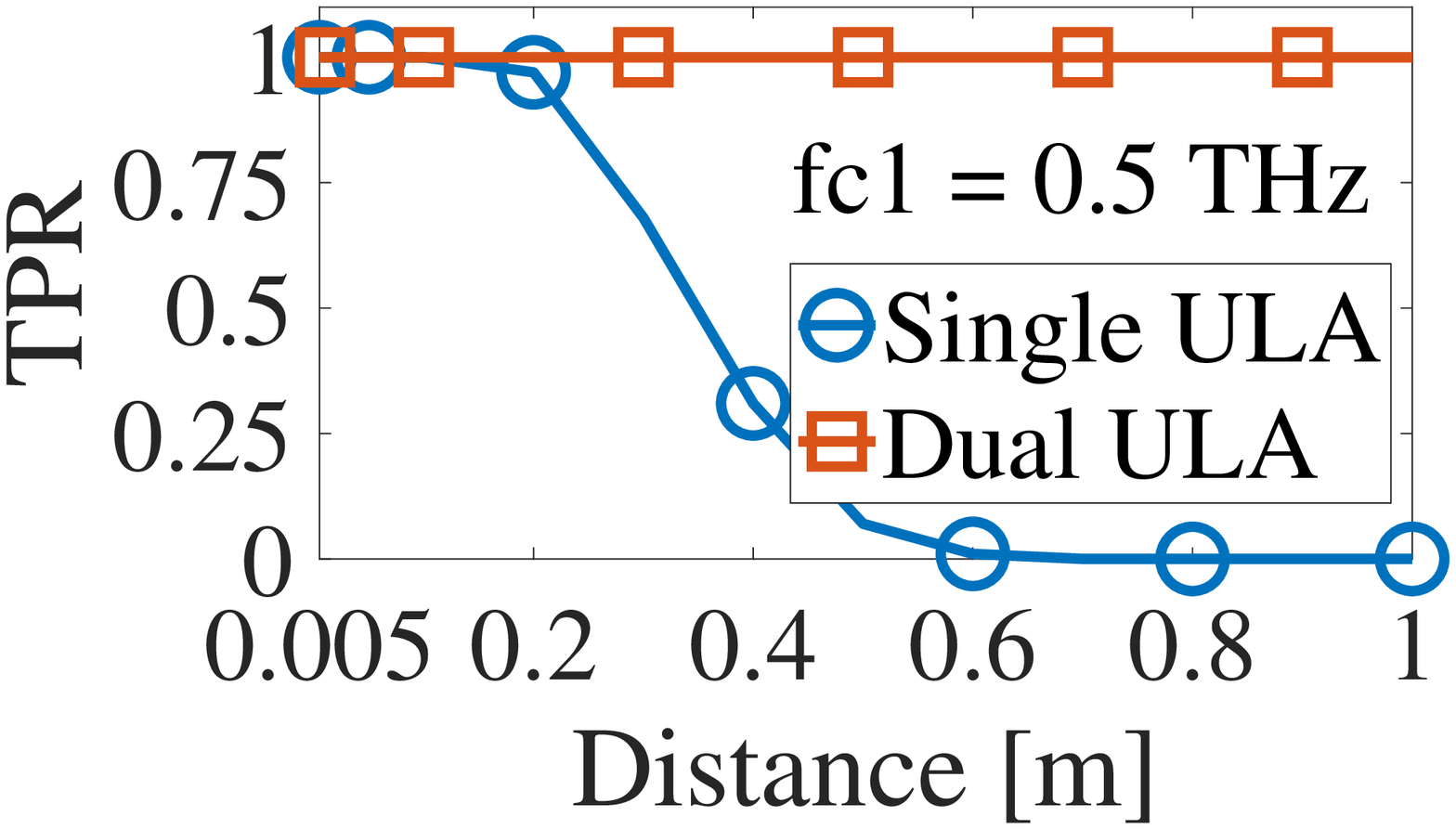}
	}
	\hspace{-7mm}
	\vspace{-2mm}
	\subfigure{
		\includegraphics[width=0.34\columnwidth,height=2.5cm]{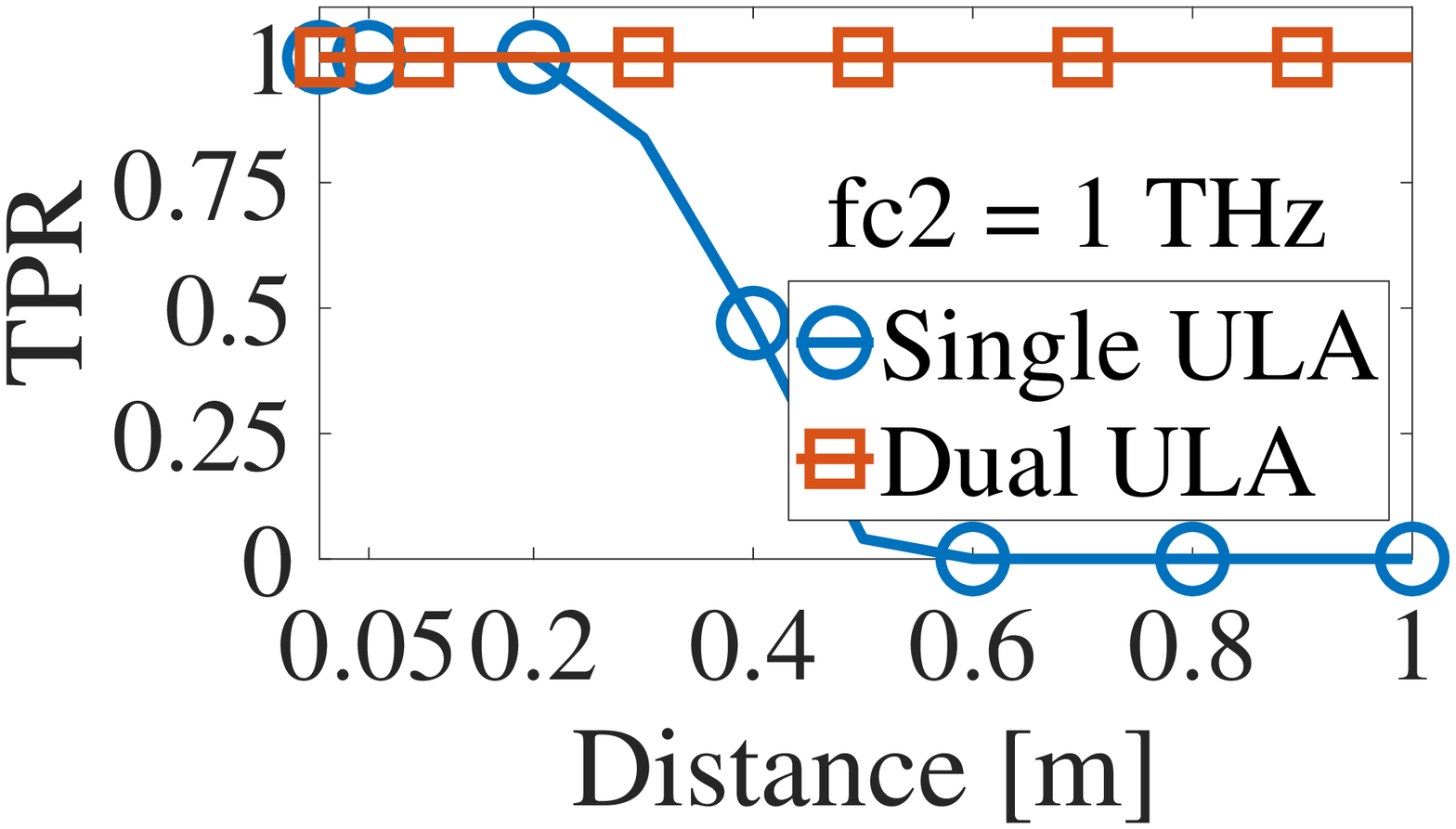}
	}
	\hspace{-7mm}
	\vspace{-2mm}
	\subfigure{
		\includegraphics[width=0.34\columnwidth,height=2.5cm]{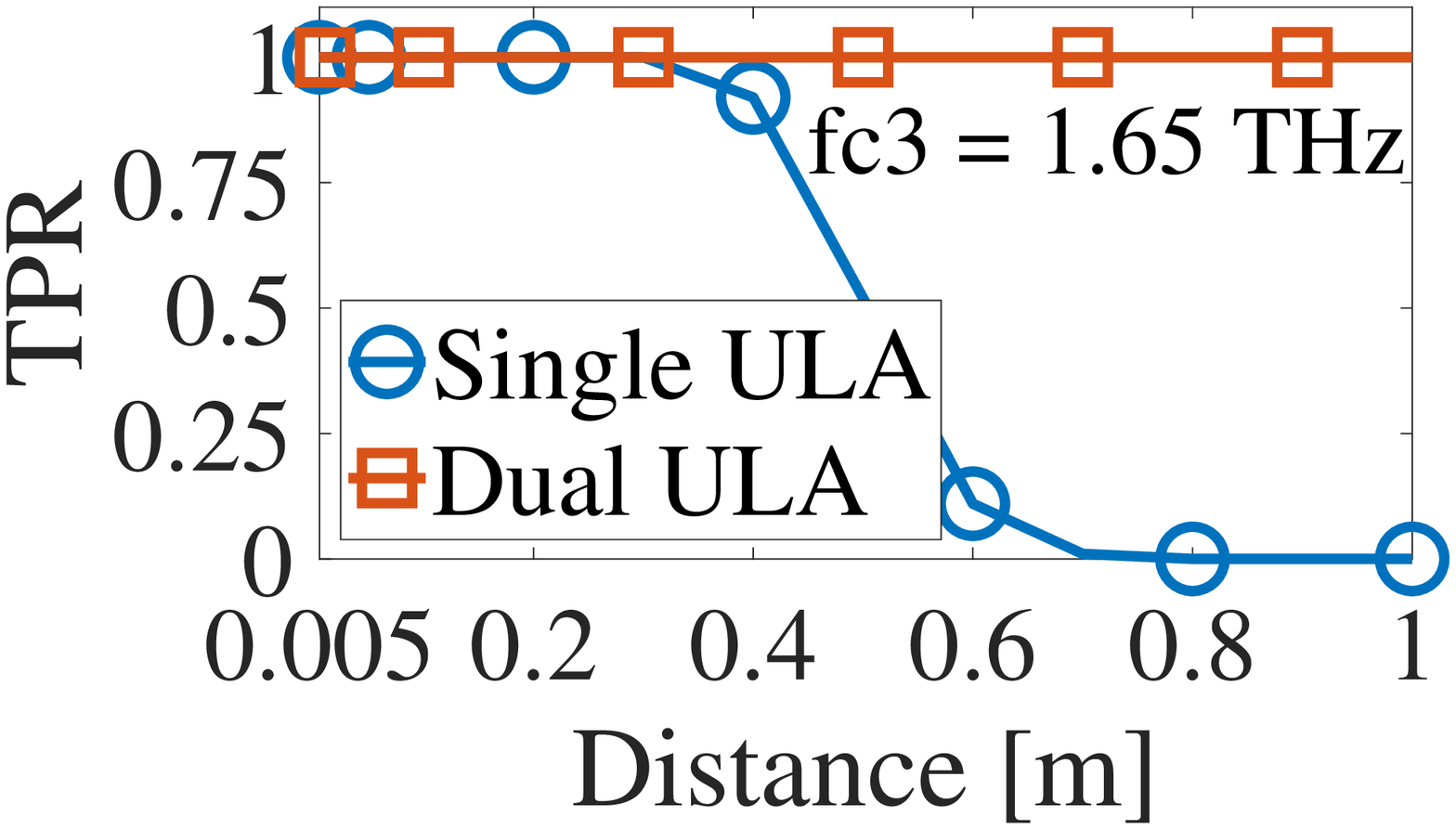}
	}
	\vspace{-2mm}
	\\
	\subfigure{
		\includegraphics[width=0.34\columnwidth,height=2.5cm]{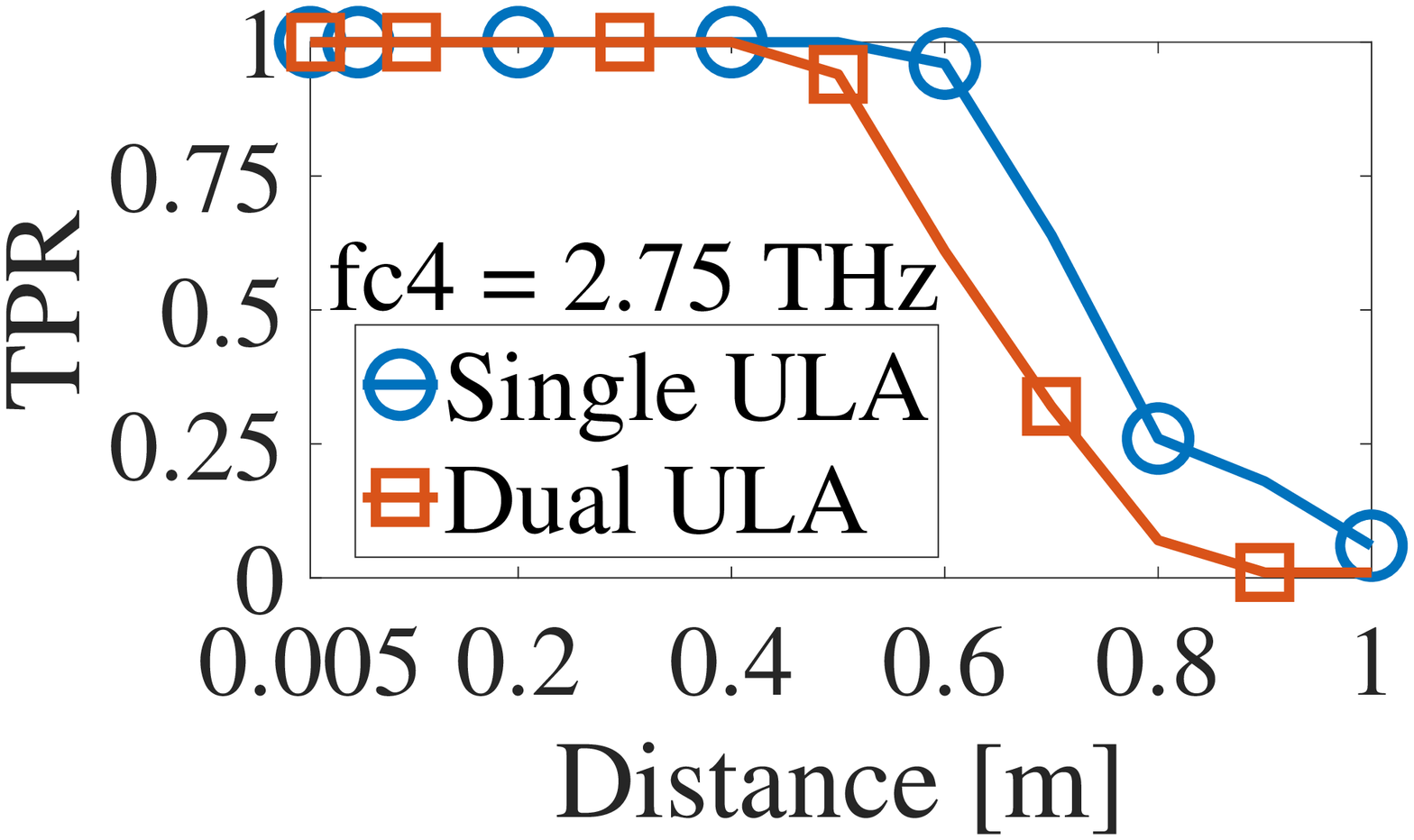}
	}
	\hspace{-7mm}
	\vspace{-0.75mm}
	\subfigure{
		\includegraphics[width=0.34\columnwidth,height=2.5cm]{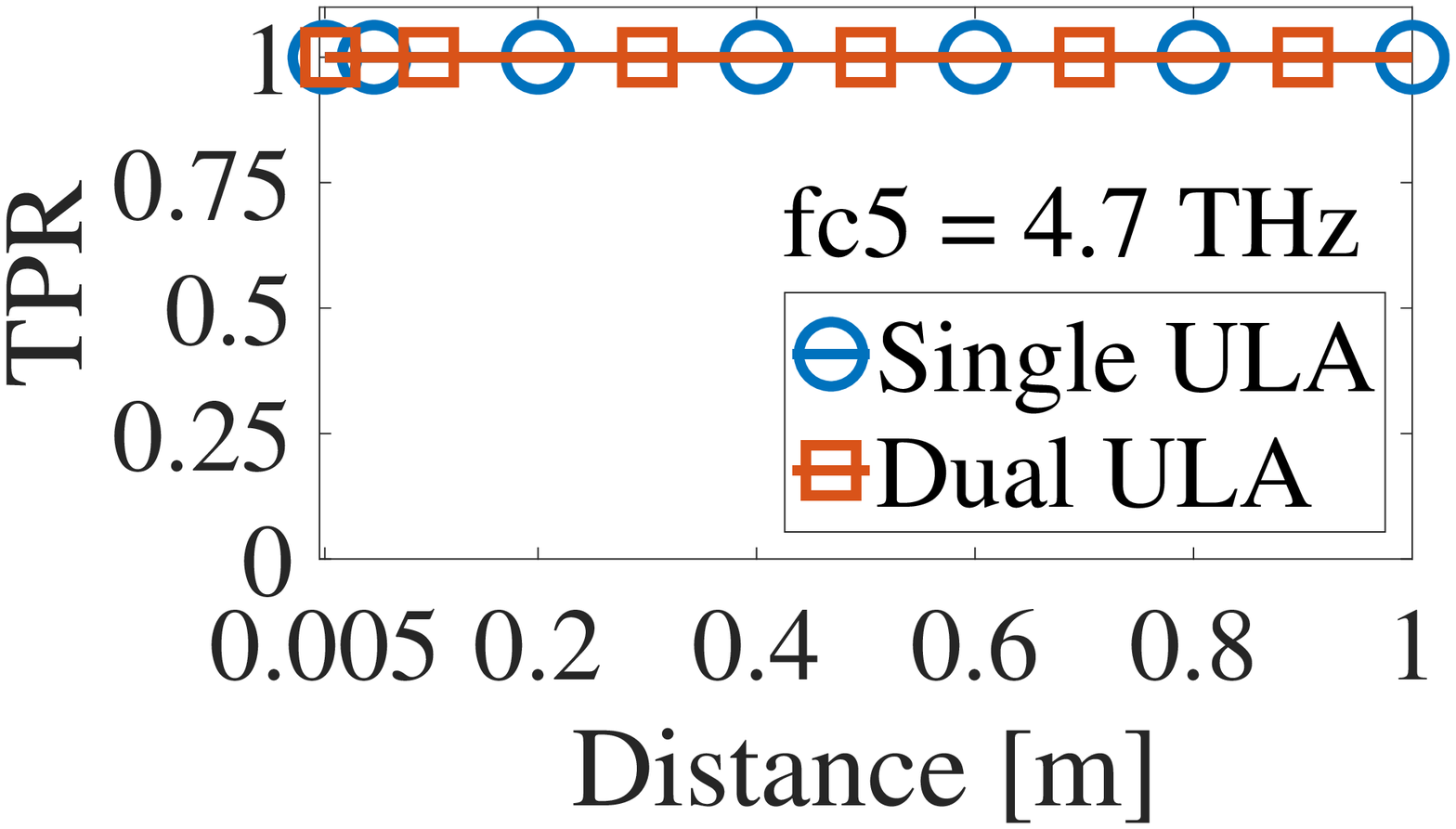}
	}
	\hspace{-7mm}
	\vspace{-0.75mm}
	\subfigure{
		\includegraphics[width=0.34\columnwidth,height=2.5cm]{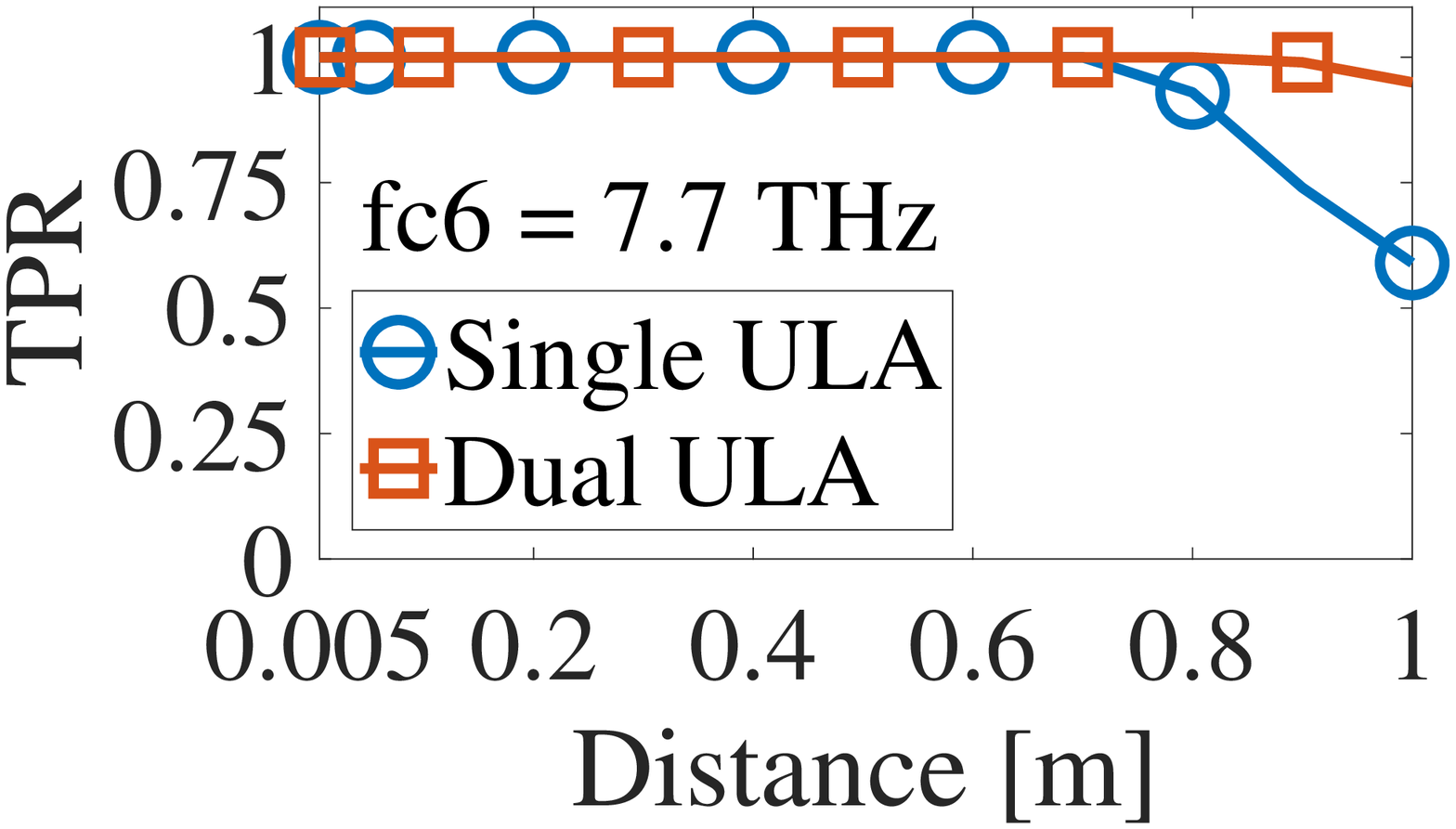}
	}
	\vspace{-5mm}
	\caption{Comparison of TPR for event classification for Single and Dual ULA.}
	\label{fig:SigDuPD}
	\vspace{-5mm}
\end{figure}

\vspace{-1mm}
RMSE for frequency estimation and TPR for event classification as a function of distance are investigated in Figs. \ref{fig:RMFRF05}, \ref{fig:RMFRF1} and \ref{fig:SigDuPD}. We observe that for very short distances below 2 cm, there is hardly any difference between single and dual ULAs and between frequencies. For distances beyond 2cm, dual ULA clearly outperforms single ULA for lower frequencies, but for higher frequencies, the differences diminish. For single ULA, TPR for the three lower frequencies plummets to zero after 0.5 m, which means that single ULA cannot support many event types.  For 1 meter distance, the confusion matrix in Table \ref{tab:Table3} captures the fact that, the single ULA erroneously classifies many lower frequencies as adjacent frequencies, which drastically reduces TPR. Excluding center frequency $f_{c_{4}}$, dual ULA provides overall TPR of 98.8\% while single ULA achieves a TPR of only 30.4\%.
\begin{table}[t]
	\centering
	\caption{Confusion matrix for center frequency estimation. Energy of sixth order Gaussian pulse is $1\:aJ$ and path length is $1\:m$}
	\vspace{-1mm}
	\label{tab:Table3}
	\resizebox{\columnwidth}{0.14\columnwidth}{\begin{tabular}{|c|c|c|c|c|c|c|c|c|c|c|c|c|c|}
			\hline
			\multicolumn{2}{|c|}{\multirow{3}{*}{}} & \multicolumn{6}{c|}{Single ULA} & \multicolumn{6}{c|}{Dual ULA} \\ \cline{3-14} 
			\multicolumn{2}{|c|}{} & \multicolumn{6}{c|}{True Frequencies} & \multicolumn{6}{c|}{True Frequencies} \\ \cline{3-14} 
			\multicolumn{2}{|c|}{} & $f_{c_{1}}$ & $f_{c_{2}}$ & $f_{c_{3}}$ & $f_{c_{4}}$ & $f_{c_{5}}$ & $f_{c_{6}}$ & $f_{c_{1}}$ & $f_{c_{2}}$ & $f_{c_{3}}$ & $f_{c_{4}}$ & $f_{c_{5}}$ & $f_{c_{6}}$ \\ \hline
			\multirow{6}{*}{\rotatebox{90}{\parbox{1cm}{Estimated\\Frequency}}} 
			&$f_{c_{1}}$&\textbf{ 0} & 0 & 0 & 0 & 0 & 0 & \textbf{98} & 0 & 0 & 0 & 0 & 0 \\ \cline{2-14} 
			& $f_{c_{2}}$ &1 &\textbf{0} & 0 & 0 & 0 & 0 & 2 & \textbf{100} & 0 & 0 & 0 & 0 \\ \cline{2-14} 
			& $f_{c_{3}}$ &36 & 3 & \textbf{0} & 0 & 0 & 0 & 0 & 0 & \textbf{100} & 0 & 0 & 0 \\ \cline{2-14} 
			& $f_{c_{4}}$ & 58 & 90 & 65& \textbf{3} & 0 & 0 & 0 & 0 & 0 & \textbf{0} & 0 & 0 \\ \cline{2-14} 
			& $f_{c_{5}}$ & 5 & 7 & 35 & 97 & \textbf{100} & 48 & 0 & 0 & 0 & 100 & \textbf{100} & 4 \\ \cline{2-14} 
			& $f_{c_{6}}$ & 0 & 0 & 0 & 0 & 0 & \textbf{52} & 0 & 0 & 0 & 0 & 0 & \textbf{96} \\ \hline
			\multicolumn{2}{|c|}{TPR} & 0 & 0 & 0 & 0.03 & 1 & 0.52 & 0.98 & 1 & 1 & 0 & 1 & 0.96 \\ \hline
			\multicolumn{2}{|c|}{\parbox{1.85cm}{Overall TPR\\(Excluding $f_{c_{4}}$)}} & \multicolumn{6}{c|}{\textbf{0.304}} & \multicolumn{6}{c|}{\textbf{\text{0.988}}} \\ \hline
	\end{tabular}}
	\vspace{-6mm}
\end{table}
\vspace{-2mm}
\section{Conclusion and Future Work}
\vspace{-1.1mm}
In this paper, we propose a framework to locate and classify different events detected by nanosensor in nano IoT using single higher order Gaussian pulse. Nanosensors are localized by estimating their DOA using wideband MUSIC algorithm. Different events sensed by nanosensors are encoded in different center frequencies with non overlapping half power bandwidth. The spectral centroid is used by the base station to classify events conveyed by nanosensors. Our simulations show that dual ULA provides good localization and event classification accuracy as compared to using single ULA for the entire terahertz band. Using dual ULA it is possible to localize and classify five different events with an accuracy of  $1.58^{\circ}$ and 98.8\% from a single pulse transmitted by a nanosensor device. In future, event classification and localization will be investigated when multiple nanosensor devices detect events and transmit pulses at the same time. Further, the classification of the center frequency of pulses will also be studied using asynchronous receivers.
\vspace{-3mm}
\section*{Acknowledgment}
\vspace{-1mm}
The proposed work is supported by SERB, GOI under order no. SB/S3/EECE/210/2016

\end{document}